\documentclass[traditabstract]{aa}
\usepackage{graphicx}
\usepackage{rotating,subfigure,amssymb,afterpage}
\usepackage{txfonts}
\usepackage{natbib}
\begin{document}

   \title{The Cosmic Large-Scale Structure in X-rays (CLASSIX) \\
   Cluster Survey III:
The Perseus-Pisces supercluster and the Southern Great Wall as traced by X-ray
luminous galaxy clusters
\thanks{
   Based on observations at the European Southern Observatory La Silla,
   Chile, and the German-Spanish Observatory at Calar Alto}}

   \author{Hans B\"ohringer\inst{1,2}, Gayoung Chon\inst{1}, Joachim Tr\"umper\inst{2}} 

   \offprints{H. B\"ohringer, hxb@mpe.mpg.de}

   \institute{$^1$ Universit\"ats-Sternwarte M\"unchen, Fakult\"at f\"ur Physik,
                  Ludwig-Maximilians-Universit\"at M\"unchen,
                  Scheinerstr. 1, 81679 M\"unchen, Germany.\\
              $^2$ Max-Planck-Institut f\"ur extraterrestrische Physik,
                   D-85748 Garching, Germany.
}

   \date{Submitted 23/03/2021}

\abstract{The Perseus-Pisces supercluster is known as one of the largest structures in the nearby
Universe that has been charted by the galaxy and galaxy
cluster distributions. For the latter mostly clusters from the Abell catalogue have been
used. Here we take a new approach to a quantitative characterisation of the Perseus-Pisces
supercluster using a statistically complete sample of X-ray luminous galaxy groups and clusters
from our {\sf CLASSIX} galaxy cluster redshift survey. We used a friends-of-friends technique to
construct the supercluster membership. We also studied  the structure of the Southern Great Wall,
which merges with the Perseus-Pisces supercluster with a slightly 
increased friends-of-friends linking length. In this work we discuss the geometric structure of the superclusters,
compare the X-ray luminosity distribution of the members with that of the surroundings,
and provide an estimate of the supercluster mass. These results establish Perseus-Pisces
as the largest superstructure in the Universe at redshifts $z \le 0.03$.
With the new data this supercluster extends through the zone of avoidance,
which has also been indicated by some studies of the galaxy distribution 
by means of HI observations.  
We investigated whether the shapes of the member groups and clusters in X-rays
are aligned with the major axis of the supercluster.
We find no evidence for a pronounced alignment, except for the ellipticities
of Perseus and AWM7, which are aligned with the separation vector of the two
systems and weakly with the supercluster.
}


\authorrunning{B\"ohringer et al.}
\titlerunning{Perseus-Pisces supercluster}
   \maketitle
%

\section{Introduction}

The large-scale structure of the Universe is often described as
the cosmic web traced by galaxies and galaxy clusters. In the hierarchy 
of structures  superclusters (SCs) are the next largest units above galaxy
clusters (e.g. {\citealt{Bah1988, Ein2007}). 
The term superclusters is not very well defined, and is applied to
a variety of structures with a large range of sizes and overdensities.
We  therefore introduced a  clearer definition of superclusters
at higher overdensities, comprising those supercluster mass distributions
that will collapse in the future in a Lambda cold dark matter ($\Lambda$CDM) Universe, calling them
superstes-clusters \citep{Cho2015}\footnote{Superstes is Latin for 
survivor, and describes the fate of these structures that have survived a dispersion
in a Universe with accelerated expansion.}. Structures larger than this
will be transient phenomena in the long run and we refer to them here as superclusters. 
They are still in the regime of non-linear structures,
as reflected in their flattened or filamentary shapes and overdensities
well exceeding unity. On scales larger than $\sim 300$ Mpc, the density
fluctuations approach a Gaussian distribution and
the statistics can be fairly well described by the 
matter distribution power spectrum. In this paper
and the other publications in this series we study the superclusters
in the nearby Universe using galaxy clusters as probes.

In this paper  we characterise the Perseus-Pisces SC
and the neighbouring Southern Great Wall (SGW). The Perseus-Pisces SC
was found to be one of the largest SCs in the early
exploration of the cosmic web (e.g. \citealt{Joe1978,Gre1981,Chi1983,Bat1985}). 
\citet{Joe1978} found the Perseus-Pisces SC to be the most prominent
structure in the Universe out to a recession velocity of 15 000 km s$^{-1}$ 
($\sim 214~h_{70}^{-1}$ Mpc) among other 
chains of clusters in the constellations Coma, Hercules, and Fornax. They
describe the Perseus-Pisces SC to cover about 45 degrees of the sky
comprising the clusters Abell 426 (Perseus), A262, A347 and the galaxy groups 
associated with the giant galaxies NGC315, NGC 383, NGC 507, and NGC 1129.
Based on a total luminosity of about $10^{13}~h_{50}^{-2}$ L$_{\odot}$ they attribute a mass
of about $2 \times 10^{16}~h_{50}^{-1}$ M$_{\odot}$ to this SC and find that it  has an
overdensity compared to the cosmic mean of about a factor of 3. They remark that all
member clusters are elongated along the main ridge of the chain.

\citet{Gre1981} described the Perseus-Pisces
SC as a concentration of clusters on an area of about $40 \times 12$ degrees
on the sky, occupying a volume of about $2 \times 10^{5}~h^{-3}$ Mpc$^3$
at a mean redshift of $z = 0.0178$. They associate similar members with the 
SC, as did Joeveer \& Einasto. \citet{Bat1985} isolated a 
larger structure in their study out to $z=0.08$, named the Perseus-Pegasus 
filament, with an extent of 300 Mpc. This structure is much larger than the 
Perseus-Pisces SC considered here; it  extends into the constellation of Pegasus
and Aquarius and reaches  higher redshifts than the structures described 
above. More recent works (e.g. \citealt{Hau1987,Cha1990,Ram2016,Kra2018}) based
on radio observations of HI in galaxies  discuss various extensions of
the Perseus-Pisces SC, also into the zone of avoidance (ZoA), which is 
discussed in section 5.

In this study we used X-ray luminous galaxy clusters for the large-scale structure mapping.
Since clusters form from the largest peaks in the initially random Gaussian density fluctuation 
field, their density distribution can be statistically closely related 
to the matter density distribution (e.g. \citealt{Bar1986}). 
Using X-ray luminosity to characterise the clusters has two advantages.
The X-ray emission ensures that  the systems are   tight three-dimensional mass 
concentrations, and the X-ray luminosity provides
a good estimate of the object's mass.

Cosmic
structure formation theory has shown that the ratio of the cluster density
fluctuation amplitude is biased with respect to the matter density fluctuations
in the sense that the cluster density fluctuations follow the matter
density fluctuations with a larger amplitude. This bias is practically 
scale-independent
(e.g. \citealt{Kai1986,Mo1996,She1999,Tin2010}). 
The  bias of the cluster density fluctuation amplitude is an enhancement,
which makes clusters sensitive tracers of the matter distribution.

For our study we used the large, highly complete sample of X-ray luminous galaxy
clusters from the Cosmic Large-Scale Structure in X-rays ({\sf CLASSIX})  galaxy 
cluster survey. The survey is flux-limited, which 
provides an X-ray luminosity-limited (and closely mass-limited) cluster sample
in each redshift shell. We have already applied 
{\sf CLASSIX} to successful studies of the cosmic large-scale structure, as
explained in section 2, and the present study can build on this experience.

   \begin{table*}
      \caption{Galaxy group and cluster members of the Perseus-Pisces SC. The flux is in units of $10^{-12}$ 
erg s$^{-1}$ cm$^{-2}$ in the 0.1 - 2.4 keV band, and the error in the following column is in per cent. The X-ray 
luminosity, $L_X$, is in units of $10^{44}$ erg s$^{-1}$ for 0.1 to 2.4 keV within $r_{500}$; $m_{200}$ is 
the cluster mass estimated from the $L_X$-mass relation within $r_{200}$; $r_{out}$ is the radius out to
which the X-ray luminosity is detected in the RASS; and $n_H$ is the interstellar column density 
in the line of sight in units of $10^{20}$ cm$^{-2}$. $R$ indicates to which part of
the Perseus-Pisces SC the member belongs: C = core, D = other members $z \le 0.025$, E = members with
$z = 0.025 - 0.03$, and F = members at $z > 0.03$. In the last column we provide alternative names
of the systems; groups are often designated by the name of the central dominant galaxy.}
         \label{Tempx}
      \[
         \begin{array}{lrrrrrrrrrrl}
            \hline
            \noalign{\smallskip}
{\rm name}&{\rm RA}&{\rm DEC}&{\rm redshift}&{\rm flux}& {\rm err.}&L_X&m_{200}&r_{out}&n_H & R  &{\rm alt. name} \\
            \noalign{\smallskip}
            \hline
            \noalign{\smallskip}
{\rm RXCJ2332.5+2355}& 353.1303 &  23.9266 & 0.0173 &   1.6108 &  17.50 &   0.0143 &   0.266 &   7.5 &   4.4&D&{\rm UGC~12655}\\
{\rm RXCJ0015.5+1720}&   3.8764 &  17.3350 & 0.0181 &   1.9179 &  35.80 &   0.0181 &   0.308 &   8.0 &   4.0&D&{\rm NGC~57}\\
{\rm RXCJ0018.3+3003}&   4.5984 &  30.0630 & 0.0223 &   2.5900 &  20.10 &   0.0374 &   0.484 &   7.5 &   5.5&D&{\rm NGC~71}\\
{\rm RXCJ0021.0+2216}&   5.2628 &  22.2754 & 0.0193 &   5.1874 &  25.50 &   0.0464 &   0.554 &  16.0 &   4.2&D&{\rm SRGb063}^{a)},{\rm ~PPS~62}^{b)}\\
{\rm RXCJ0107.2+3224}&  16.8206 &  32.4067 & 0.0174 &   6.6932 &  10.30 &   0.0512 &   0.590 &  14.5 &   5.2&C&{\rm NGC~383}\\
{\rm RXCJ0110.9+3308}&  17.7425 &  33.1486 & 0.0177 &   2.2140 &  20.00 &   0.0277 &   0.404 &   5.0 &   5.9&C&{\rm NGC~410}\\
{\rm RXCJ0123.1+3327}&  20.7971 &  33.4612 & 0.0147 &   6.0909 &  18.00 &   0.0416 &   0.519 &   9.0 &   5.3&C&{\rm NGC~499}\\
{\rm RXCJ0123.6+3315}&  20.9041 &  33.2516 & 0.0166 &  17.1433 &  15.00 &   0.1523 &   1.159 &  10.0 &   5.3&C&{\rm NGC~507}\\
{\rm RXCJ0152.7+3609}&  28.1942 &  36.1509 & 0.0161 &  71.6749 &   4.50 &   0.4227 &   2.185 &  38.0 &   5.5&C&{\rm A~262}\\
{\rm RXCJ0200.2+3126}&  30.0704 &  31.4346 & 0.0167 &   4.0825 &  16.00 &   0.0308 &   0.430 &  11.0 &   5.5&C&{\rm NGC~777}\\
{\rm RXCJ0222.7+4301}&  35.6965 &  43.0182 & 0.0212 &   7.0947 &  20.80 &   0.0910 &   0.841 &   9.0 &   9.1&C&{\rm UGC~1841}\\
{\rm RXCJ0238.7+4138}&  39.6977 &  41.6478 & 0.0153 &   2.3705 &  21.20 &   0.0130 &   0.253 &  14.5 &   7.6&C&{\rm NGC~996}\\
{\rm RXCJ0249.5+4658}&  42.3899 &  46.9768 & 0.0274 &   2.3755 &  19.40 &   0.0433 &   0.530 &  11.5 &  13.4&E&{\rm IC~0256/0257}\\
{\rm RXCJ0254.4+4134}&  43.6214 &  41.5764 & 0.0172 & 110.5048 &   2.20 &   0.9477 &   3.601 &  17.5 &   9.2&C&{\rm AWM7~(NGC1129)}\\
{\rm RXCJ0300.7+4428}&  45.1791 &  44.4675 & 0.0302 &  41.8408 &   4.00 &   0.9475 &   3.578 &  17.5 &  14.9&F&{\rm CIZAJ0300.7+4427}\\ 
{\rm RXCJ0309.9+4207}&  47.4981 &  42.1255 & 0.0300 &   1.4685 &  19.70 &   0.0316 &   0.434 &  11.0 &  12.7&E&{\rm UGC~2562}\\
{\rm RXCJ0310.3+4250}&  47.5755 &  42.8488 & 0.0314 &   2.6345 &  19.00 &   0.0668 &   0.681 &   9.0 &  16.2&F&{\rm }\\ 
{\rm RXCJ0319.7+4130}&  49.9498 &  41.5149 & 0.0179 & 688.3633 &   0.90 &   7.7309 &  13.227 &  17.5 &  15.7&C&{\rm A~426~(Perseus)}\\
{\rm RXCJ0348.1+4212}&  57.0337 &  42.2145 & 0.0174 &   3.3776 &  17.20 &   0.0324 &   0.443 &   7.5 &  23.7&C&{\rm MCG~+07-08-033}\\
{\rm RXCJ0421.8+3607}&  65.4689 &  36.1200 & 0.0208 &  11.6173 &  10.10 &   0.1279 &   1.038 &  14.5 &  26.2&C&{\rm UGC~3021}\\
{\rm RXCJ0450.0+4501}&  72.5146 &  45.0242 & 0.0210 &  69.6988 &   5.50 &   0.8256 &   3.300 &  17.5 &  67.9&C&{\rm 3C~129}\\
{\rm RXCJ0547.2+5052}&  86.8160 &  50.8786 & 0.0263 &   9.1313 &  11.70 &   0.1490 &   1.139 &  17.0 &  17.8&E&{\rm UGC~3355}\\
            \noalign{\smallskip}
            \hline
            \noalign{\smallskip}
         \end{array}
      \]
{Notes:$^{a)}$ Group described in \citet{Mah2000}, $^{b)}$ system in the Perseus-Pisces Group Survey of \citet{Tra1998}} 
\label{tab1}
   \end{table*}

The paper has the following structure. In section 2 we describe 
the {\sf CLASSIX} galaxy cluster survey. Section 3 deals with methodological aspects. 
The results of our analysis is presented in section 4. Section 5
provides a discussion. A summary and conclusions are given in section 6.
In the Appendix we provide combined X-ray and optical images of the members of the
Perseus-Pisces SC and the SGW, investigate the alignment of cluster shapes 
with the SC main axis in more detail,
and discuss the morphology of one of the most compact groups in our sample,
NGC 410.
For physical properties that depend on distance we use the following
cosmological parameters: $H_0 = 70$ km s$^{-1}$ Mpc$^{-1}$,
$\Omega_m = 0.3$, and a spatially flat metric. For the cosmographical analysis we
use Supergalactic coordinates, defined by the location of the
Supergalactic north pole at $l_{II} = 47.3700^o$ and $b_{II} = 6.3200^o$,
as established by De Vaucouleurs et al. in the Third Catalogue 
of Bright Galaxies (1991, see also \citet{Lah2000}). For the X-ray luminosities
quoted in the following we use the ROSAT band, $0.1 - 2.4$ keV.

\section{The CLASSIX Galaxy Cluster Survey}

The {\sf CLASSIX} galaxy cluster catalogue provides an ideal database
for this study since the cluster density is high enough to provide
a meaningful dense sampling of the large-scale matter distribution
and the selection function is well understood, allowing an unbiased
mapping of this distribution. {\sf CLASSIX} is the combination of our 
surveys in the southern sky,  {\sf REFLEX II}
\citep{Boe2013}, and the northern hemisphere, {\sf NORAS II}
\citep{Boe2017}. The total coverage is 8.26 ster of the sky
at galactic latitudes $|b_{II}| \ge 20^o$l. 
An extension of {\sf CLASSIX} also includes lower galactic 
latitudes, part of the ZoA, in a region 
restricted to interstellar hydrogen column density $n_H \le 2.5 \times 10^{21}$ cm$^{-2}$.
At higher column density X-rays are strongly absorbed in the ROSAT energy band,
and  the sky usually has a high stellar density, making the detection
of clusters in the optical extremely difficult. The values for the
interstellar hydrogen column density are taken from the 21cm survey of 
\citet{Dic1990}~\footnote{We  compared the
interstellar hydrogen column density compilation by \citet{Dic1990}
with the more recent data set of the 
Bonn-Leiden-Argentine 21cm survey \citep{Kal2005},
and found that the differences relevant for us are   at most one per cent.
Because our survey was constructed with a flux cut based
on the Dickey \& Lockman results, we decided to keep the older hydrogen column density
values for consistency.}.
This area adds another 2.56 ster and increases the sky coverage to 86.2\% 
of the sky. The completeness of the cluster detection in this area 
is not as high as for {\sf REFLEX} and {\sf NORAS,} and 
follow-up to obtain redshifts for this part of the survey is still incomplete.
The statistical properties of the cluster distribution in the ZoA are 
therefore somewhat qualitative. In addition, three known X-ray luminous
clusters in the region with $n_H > 2.5 \times 10^{21}$ cm$^{-2}$, which fulfil
the X-ray selection parameters, were added to the cluster sample. 

The {\sf CLASSIX} galaxy cluster survey and its extension is based on 
the X-ray detection of galaxy clusters in the ROSAT All-Sky Survey
(RASS, \citet{Tru1993,Vog1999}. The construction of the cluster survey 
and the survey selection function  as well as tests of the completeness
are described in \citet{Boe2013,Boe2017}. In summary, the 
nominal unabsorbed flux limit for the galaxy cluster detection in the RASS is
$1.8 \times 10^{-12}$ erg s$^{-1}$ cm$^{-2}$ in the
0.1 - 2.4 keV energy band, and we require a minimum count of detected source photons of 20. 
Under these conditions the nominal flux limit quoted above is reached in about
80\% of the survey. In regions with lower exposure and higher interstellar
absorption the flux limit is accordingly higher 
(see Fig.\ 11 in \citealt{Boe2013} and Fig.\ 5 in \citealt{Boe2017}), 
and this effect is well modelled and
taken into account in the survey selection function.

We already applied the {\sf REFLEX I}  
\citep{Boe2004} and  {\sf REFLEX II} surveys
to study the cosmic large-scale matter distribution, for example
through  the correlation function \citep{Col2000}, 
the power spectrum \citep{Sch2001,Sch2002,Sch2003a,Sch2003b,
Bal2011,Bal2012}, 
and Minkowski functionals, \citep{Ker2001}.
We  compiled a catalogue of superstes-clusters and studied
their properties \citep{Cho2013,Cho2014}. 
We  find from the cluster distribution that the local Universe
has a lower matter density inside a radius of $\sim 100 - 170$ Mpc
than the cosmic mean on larger scales \citep{Boe2015,Boe2020},
and we  show that on a scale of 
$\sim 100$ Mpc the matter distribution is strongly segregated onto the 
Supergalactic plane \citep{Boe2021}. 

The two most essential physical parameters of the clusters are
their X-ray luminosity and mass. X-ray luminosities in the 0.1 to
2.4 keV energy band were  derived within a cluster radius of 
$r_{500}$ \footnote{$r_{500}$ is the radius where the average
mass density inside reaches a value of 500 times the critical density
of the Universe at the epoch of observation.}. To estimate the cluster
mass and temperature from the observed X-ray luminosity, we use the 
scaling relations described in \citet{Pra2009}. They were
determined from a representative cluster sub-sample of our survey
called {\sf REXCESS} \citep{Boe2007}. 

The survey selection function determined as a function
of the sky position 
and as a function of redshift, an important ingredient for our study, is documented
in  \citet{Boe2013} for {\sf REFLEX II} and \citet{Boe2017}
for {\sf NORAS II}, where   numerical data are also provided in the online material.
   \begin{table*}
      \caption{Galaxy group and cluster members of the Southern Great Wall. The meaning of the columns
is the same as in Table 1.}
         \label{Tempx}
      \[
         \begin{array}{lrrrrrrrrrl}
            \hline
            \noalign{\smallskip}
{\rm name}&{\rm RA}&{\rm DEC}&{\rm redshift}&{\rm flux}& {\rm err.}&L_X&m_{200}&r_{out}&n_H & {\rm alt. name} \\
            \noalign{\smallskip}
            \hline
            \noalign{\smallskip}
{\rm RXCJ0125.5+0145}&  21.3757 &   1.7623 & 0.0184 &   5.3397 &  15.00 &   0.0452 &   0.546 &  14.0 &   3.1&{\rm NGC~533}\\
{\rm RXCJ0125.6-0124}&  21.4198 &  -1.4072 & 0.0180 &  11.7510 &  14.10 &   0.0863 &   0.816 &  24.0 &   4.1&{\rm A~194}\\
{\rm RXCJ0149.2+1303}&  27.3049 &  13.0649 & 0.0171 &   2.3451 &  20.50 &   0.0188 &   0.318 &   9.5 &   4.9&{\rm NGC~677}\\
{\rm RXCJ0156.3+0537}&  29.0924 &   5.6228 & 0.0185 &   3.0455 &  13.20 &   0.0281 &   0.407 &  10.0 &   4.3&{\rm NGC~741}\\
{\rm RXCJ0231.9+0114}&  37.9881 &   1.2445 & 0.0218 &   2.9004 &  21.10 &   0.0326 &   0.445 &  14.5 &   2.9&{\rm UGC~2005}\\
{\rm RXCJ0252.8-0116}&  43.2060 &  -1.2741 & 0.0235 &   7.4523 &  16.20 &   0.1150 &   0.971 &   9.5 &   5.3&{\rm NGC~1132}\\
{\rm RXCJ0257.6+0600}&  44.4038 &   6.0160 & 0.0243 &  17.1120 &  12.00 &   0.2435 &   1.546 &  18.0 &   9.3&{\rm A~400}\\
            \noalign{\smallskip}
            \hline
            \noalign{\smallskip}
         \end{array}
      \]
\label{tab1}
   \end{table*}

\section{Methods}


For the construction of the SCs from the cluster sample we used 
a friends-of-friends method. This can be applied straightforwardly to a homogeneous,
volume-limited survey. To account for the spatially varying survey limits in
our flux limit survey, we introduced a weighing scheme that is connected to the
estimated local cluster density. The weights were calculated
from an integration of the luminosity function, $\phi(L_X)$, as 

\begin{equation}
w_i = {\int_{L_{X_0}}^{\infty} \phi(L) dL \over \int_{L_{X_i}}^{\infty} \phi(L) dL} ~~~, 
\end{equation}

where $L_{X_0}$ is the nominal lower limit of the sample 
and $L_{X_i}$ is the lower X-ray luminosity limit that can be reached at the 
sky location and redshift of the cluster. The weighting scheme then accounts for
the missing cluster density if the local detection limit for the X-ray luminosity
is higher than $L_{X_0}$. 

In the friends-of-friends method we adopt a minimum linking 
length, $l_0$, when the nominal luminosity limit is reached. We adjust
the linking length $l_i = l_0 \times (w_i)^{1/3}$ if $L_{X_i}$ is higher 
than $L_{X_0}$. Since the linking length is calculated for each cluster at its
location, we take the average $l_i$ as the linking length 
between the two clusters in the linking process.
To introduce a higher weight for the shorter linking length, we average in the following
way: $<l> = l_0 \cdot (2/(1/w_1 + 1/w_2))^{1/3}$. In our study the linking length can increase by up to
about 40\% in the extragalactic sky, by up to 70\% in the ZoA.

For the main analysis we used a minimum linking length of $l_0 = 19$ Mpc. This corresponds
roughly to an overdensity ratio, $R_{Cl} = n_{Cl}/<n_{Cl}>$, of about a factor of 2  
compared to the mean density of clusters in the local Universe. 
This value is much lower than the value of 
$7.8$ needed for mass concentrations to collapse in the future in a $\Lambda$CDM
Universe (see the definition of superstes-clusters in Chon et al. 2015); nevertheless,
these structures are highly non-linear density perturbations.

We adopted a lower X-ray luminosity limit of $L_{{X_0}}=10^{42}$ erg  s$^{-1}$.
With this value for $L_{{X_0}}$ the survey is volume limited in most of the sky out
to a redshift of z = 0.016. The luminosity limit corresponds to
a cluster mass limit of about $2.1 \times 10^{13}$ M$_{\odot}$.  We  therefore included
less massive galaxy groups in our study. They are definitely gravitationally bound
entities, as shown by their extended X-ray emission, but they are optically often characterised
by a giant elliptical galaxy surrounded by a few smaller galaxies, which in its extreme
is called a `fossil group'. Therefore, these systems are often inconspicuous in optical surveys
and would not easily be recognised as galaxy groups, and are not included in the
characterisation of SCs. We note that this constitutes a difference
to optical studies where SC are constructed from prominent galaxy
concentrations. On the other hand, they may also include some galaxy concentrations 
that are not so tightly bound, and are therefore not X-ray luminous. 
Our study compares more closely to an analysis of cosmological N-body simulations
where groups and clusters are characterised as gravitationally bound dark matter halos.

Our analysis differs  in another aspect from the construction of SC
on the basis of the galaxy distribution. Galaxies tend to concentrate not only in clusters, 
but also in connecting filaments. Through these filamentary bridges galaxies can more
easily link large structures together, while in our survey groups are only found to
trace the most massive elongated structures. These differences should be kept in mind
in the comparison to other studies.

\section{Results}

Constructing SCs in the local Universe ($z \le 0.03$) 
in the region around the Perseus-Pisces SC with a  minimum linking 
length of $l_0 = 19$ Mpc, we find 20 {\sf CLASSIX} members defining 
the Perseus-Pisces SC. If we inspect a larger redshift interval we note that 
two additional clusters at redshifts $z = 0.0302 - 0.0314$
are linked to the Perseus-Pisces SC.
Decreasing the minimum linking length $l_0$ to 16 Mpc 
we can define a core of the Perseus-Pisces SC which
contains 13 cluster members. For the SGW we find seven  {\sf CLASSIX} cluster members. 
The member clusters and groups of the two SCs
together with their properties are listed in Tables 1 and 2, 
respectively. Appendix A1 and A5 provide optical images with X-ray surface brightness contours
of all the groups and clusters of the Perseus-Pisces SC and SGW.

\subsection{Sky distribution}

\begin{figure}[h]
   \includegraphics[width=\columnwidth]{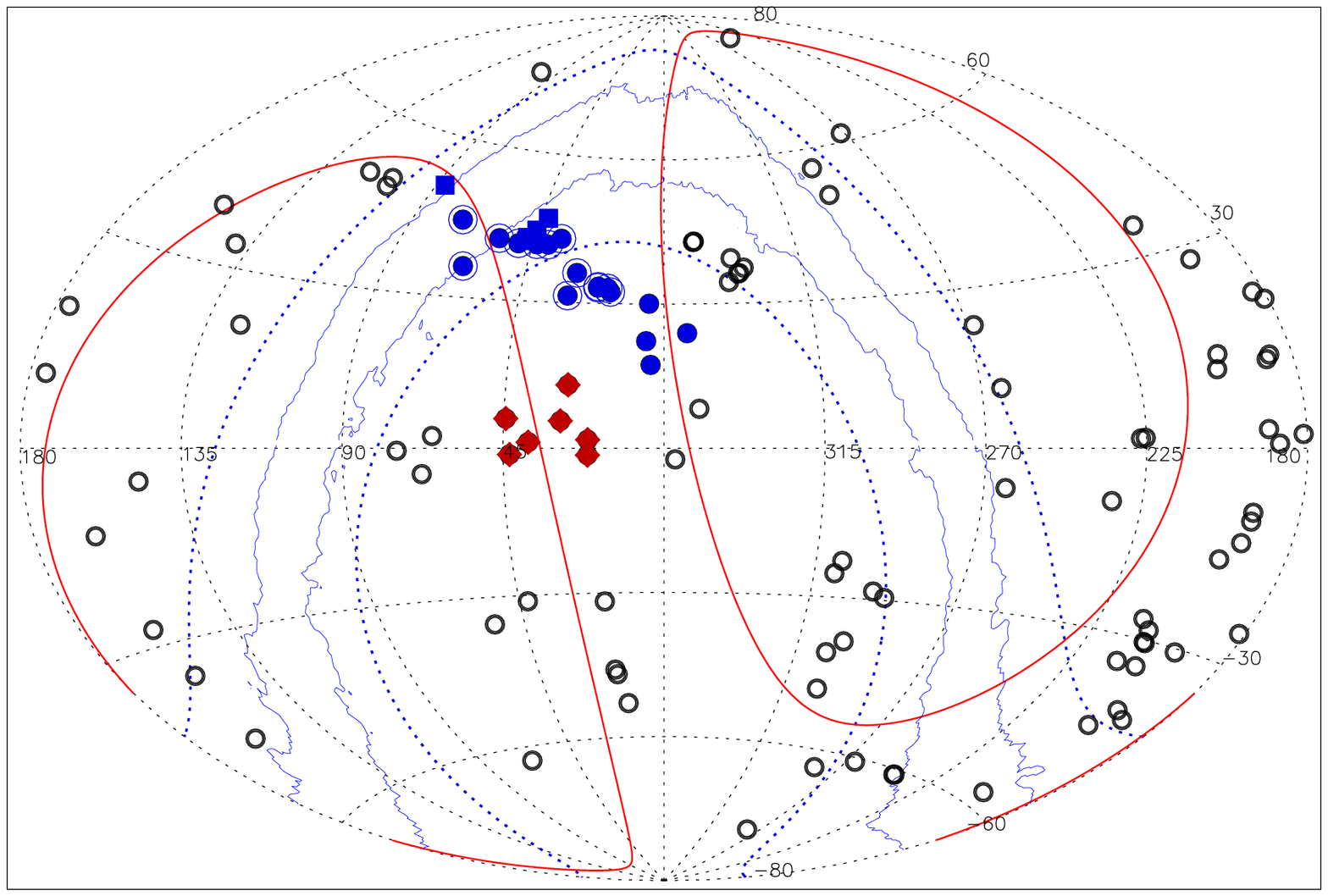}
   \includegraphics[width=\columnwidth]{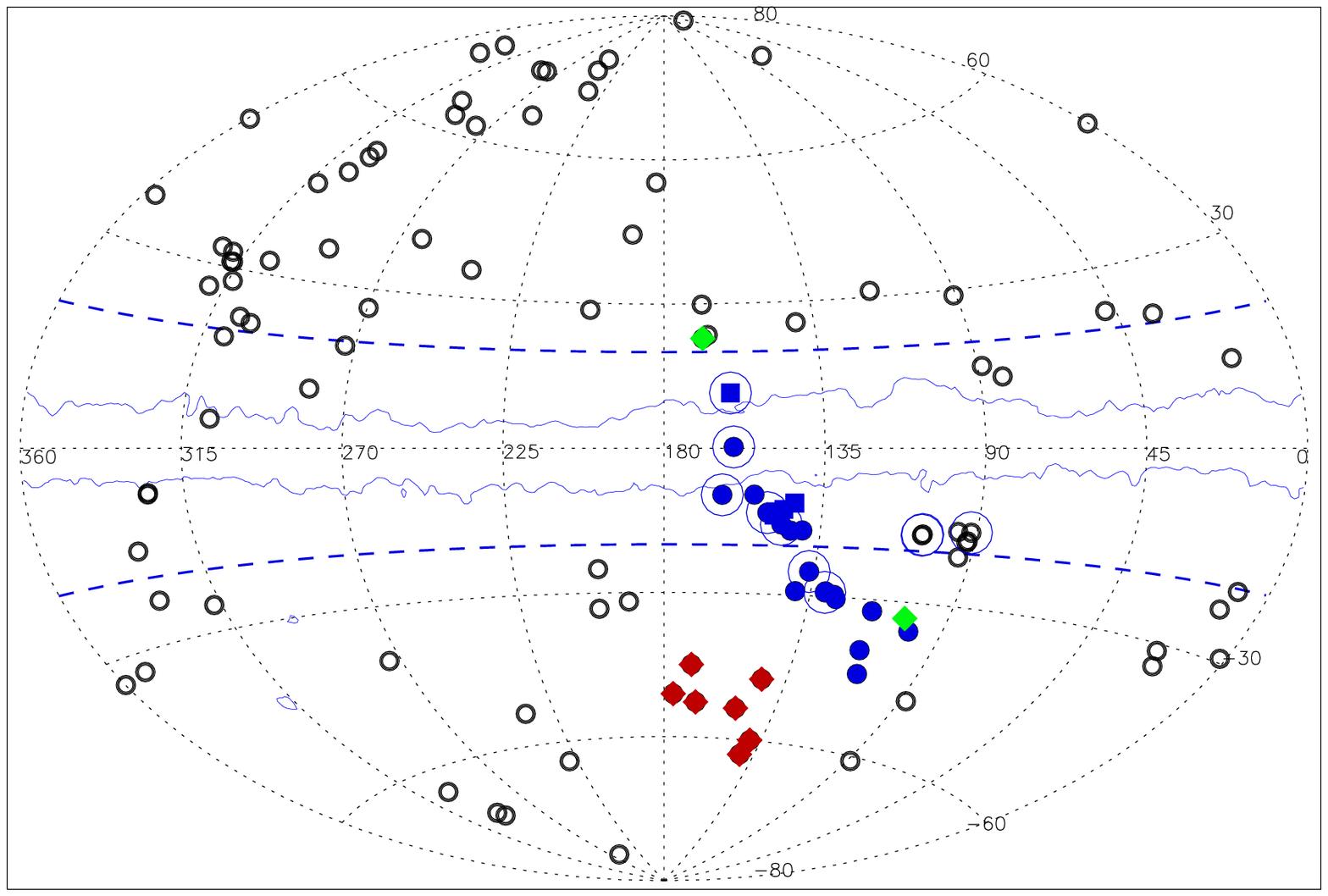}
\caption{Sky distribution of the {\sf CLASSIX} galaxy clusters in
the redshift range $z = 0 - 0.025$. The clusters associated 
with the Perseus-Pisces SC are shown as blue solid circles ($z \le 0.025$)
and blue solid squares ($z > 0.025$) and those belonging to the SGW as red 
solid diamonds.
{\bf Top:} Map in equatorial coordinates. The members of the core part of the 
Perseus-Pisces SC are shown with open circles around the circles and squares. The Galactic band
($b_{II} = \pm 20^{o}$) is shown by the blue dotted lines, the region with high hydrogen 
column density ($n_H \ge 2.5 \times 10^{21}$ cm$^{-2}$) is indicated by the solid blue lines,
and the Supergalactic band ($SGB = \pm 20^{o}$) by the red lines.
{\bf Bottom:} Map in Galactic coordinates. The larger blue open circles show
the members of the Perseus-Pisces SC found with an increased lower luminosity limit 
of $10^{43}$ erg s$^{-1}$ as explained in section 4.3. The two green points show the two
clusters A569 (positive $b_{II}$) and A2634 (negative $b_{II}$), which are part of
the discussion in section 5. 
}\label{fig1}
\end{figure}

Figure~\ref{fig1} shows the Perseus-Pisces SC and the Southern Great Wall
as they appear on the sky. The figure also shows the band
of the Milky Way ($b_{II} = \pm 20^{o}$, dotted lines), the zone of highest Galactic absorption 
(blue lines), and a region of $\pm 20^{o}$ around the Supergalactic plane (red lines). 
The three clusters at redshifts $z = 0.025 - 0.03$ and the two clusters 
at $z > 0.03$ are indicated by squares in the image.
The Perseus-Pisces SC extends over about 75 degrees on the sky. We note that it stretches
in the east across the ZoA. In Fig.~\ref{fig2},
which shows a magnification of  the SC region,  we have labelled some of the prominent clusters identified
in the figure caption. We see that it is the cluster 3C129 (number 7), close to the Galactic 
equator, that helps to bridge the SC across the ZoA. While the average
linking length for the members of the Perseus-Pisces SC is $ \sim 23$ Mpc, after weight correction
the largest linking length due to high interstellar absorption is $\sim 30$ Mpc, which bridges 
3C 129 to the galaxy group in the far  east, UGC 03355 (number 8).

In the earlier optical surveys mentioned in the
introduction the Perseus-Pisces SC ends in the east with the Perseus cluster (number 1), since
further to the east the increased extinction blocks the view. X-ray observations allow a
better view into this region, but since even our survey is incomplete in the ZoA (and
incompleteness has not been accounted for in the linking length), there may be more
SC members hidden there.

Decreasing the minimum linking length to 16 Mpc we find a denser core of the 
Perseus-Pisces SC with 13 members. These clusters are marked in the
upper panel of Fig. 1 with encircled
solid dots. The clusters belonging to the core are flagged in Table 1. Table 3
provides properties of the different selected structures of the Perseus-Pisces SC.

\begin{figure}[h]
   \includegraphics[width=\columnwidth] {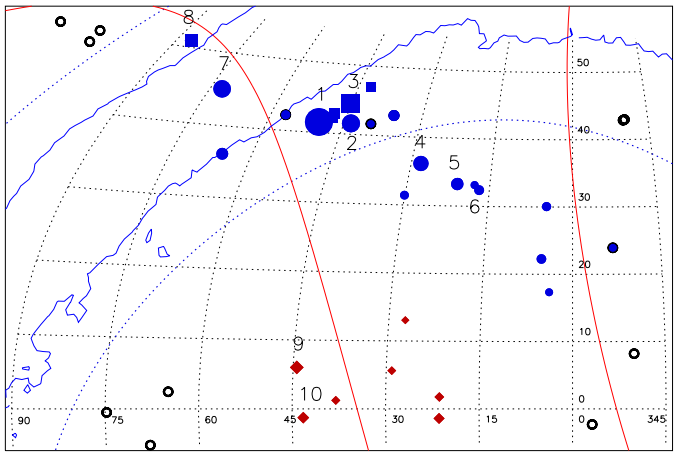}
\caption{Perseus-Pisces  SC (blue filled circles [$z\le 0.025$] and squares [$z > 0.025$]) 
and Southern Great Wall (red diamonds).
The size of the symbols indicates the estimated mass of the clusters,  as described in the text. 
The open symbols indicate clusters that are not members of these SCs. Some
massive galaxy clusters and groups are labelled: (1) Perseus cluster, (2) AWM7, 
(3) CIZAJ0300.7+4427, (4) A 262, 
(5)  NGC 499 and NGC 507 (here  blended together), (6) NGC 383, (7) 3C129, 
(8) UGC 3355, (9) A 400, (10) NGC 1132. The group UGC 2562 is hidden behind the Perseus cluster.
The lines marking the region of high interstellar hydrogen
column density, the ZoA, and the Supergalactic band are the same as in Fig.~\ref{fig1} 
}\label{fig2}
\end{figure}

In Fig.~\ref{fig2} we indicate the estimated masses of the 
SC members by the size of the symbols, with diameter scaling with the cube 
root of the estimated mass, which is the best compromise to display the mass range while minimising 
projection effects. Even so,  two clusters are hidden in projection (as noted in the
figure caption).
The mass estimates were obtained by means of the scaling relation of X-ray luminosity
and cluster mass, which has been derived from detailed XMM-Newton X-ray observations of
a representative sample of CLASSIX clusters (REXCESS), and has been parameterised as

\begin{equation}
M_{200} = 3.75 ~  L_{X,500}^{0.62}~ E(z)^{-1}~ h_{70}^{0.242} ~~,
\end{equation}

where $E(z) = H(z)^2/H_0^2$, \citep{Pra2009}. The estimated masses of 
the clusters and groups are listed in Tables 1 and 2. We note that the mass--X-ray
luminosity relation in Eq.~2 was derived for clusters with X-ray luminosities
$L_X \ge 4 \times 10^{43}$ erg s$^{-1}$.
For the mass estimates of the lower X-ray luminosity groups we rely on an extrapolation
of this relation. We expect some steepening of this relation towards lower luminosities,
due to the decreasing gas mass fraction in galaxy groups. This could be
compensated for by a larger luminosity fraction of the central cool cores found in most 
galaxy groups and by the increase in the X-ray luminosity due to the generally
higher metallicities in low mass systems. The observational results for  this in the literature
are controversial, however. This is mostly because  the mass determination in groups
from X-ray observations is difficult;  without extensively long observations only
the central regions of groups can be studied, which is not sufficient for a precise
mass estimate. Therefore, we use the extrapolation of Eq.~2. While the uncertainty
of the mass estimate for systems with $L_X \ge 4 \times 10^{43}$ erg s$^{-1}$ is about
40\%, it can be of the order of  a factor 2 for low mass systems. Apart from a small effect
on the estimated cluster masses and the sum of the cluster masses for the SC, as listed in
the tables of this paper, this has no further consequences for the construction of
the SC, except for an unsharp mass limit in the selection function.
The Perseus cluster is by far the most massive matter concentration in the Perseus-Pisces SC.
The next two largest clusters, 3C129 and AWM7, are about a factor of four less massive. 
The Southern Great Wall contains only less massive systems, where A400 with a mass of about 
$1.5 \times10^{14}$ M$_{\odot}$ is the most massive one. 

\subsection{Size and mass}

   \begin{table*}
      \caption{Properties of the Perseus-Pisces SC and the SGW
constructed with a minimum linking length of 19 Mpc. For the Perseus-Pisces SC we provide
values for three different extensions: PP = full SC, PP$_{in}$ = structure at $z \le 0.03$,
and PP$_{core}$ = core region. The columns give (from left to right)    $N_{CL}$ and $wN_{CL}$ (the number of members and the weighted
number of members, respectively); the  SC volume;  cl.~mass (the sum of the estimated 
masses of the member clusters); est.~mass and length (the estimated SC mass and its extent); 
$<z>$, $z_{min}$, and $z_{max}$  (the mean, minimum, and maximum
redshift);  the number density of clusters in the SC;  $R_{CL}$ and $R_{DM}$ (the overdensity
ratio of the cluster number density and matter density in the SC compared to the cosmic mean, 
respectively).}
         \label{T1}
      \[
         \begin{array}{lrrrrrrrrrrrr}
            \hline
            \noalign{\smallskip}
{\rm name}&{\rm N_{CL}}&{\rm wN_{CL} }&{\rm volume}&{\rm cl.~mass}& {\rm est.~mass}& {\rm length}   
        & <z> & z_{min} & z_{max} & {\rm density} & R_{CL} & R_{DM} \\
          &           &              &10^5~{\rm Mpc}^3 & 10^{14} {\rm M}_{\odot}&  10^{15} {\rm M}_{\odot}& {\rm Mpc} 
        &   &  &  & 10^{-4} {\rm Mpc}^{-3} &  &\\
            \noalign{\smallskip}
            \hline
            \noalign{\smallskip}
{\rm PP}      &  22 &  58.3 &   2.9 &   35.8 &    24.9 &   115.7 & 0.0205 & 0.0147 & 0.0314 &   2.0 &   2.8 &   2.1\\
{\rm PP}_{in} &  20 &  49.0 &    2.8 &   31.6 &   21.5 &   115.7 & 0.0195 & 0.0147 & 0.0300 &   1.8 &   2.4 &   1.9\\
{\rm PP}_{core}& 13 &  30.4 &    1.5 &   27.8 &   12.9 &    63.4 & 0.0176 & 0.0147 & 0.0210 &   2.1 &   2.9 &   2.2\\
{\rm SGW}     &   7 &  14.6 &   1.1 &    5.0 &     6.8 &    45.8 & 0.0202 & 0.0171 & 0.0243 &   1.4 &   1.9 &   1.6\\
            \noalign{\smallskip}
            \hline
            \noalign{\smallskip}
         \end{array}
      \]
\label{tab1}
   \end{table*}

The total length of the Perseus-Pisces SC, the largest separation between 
two member clusters in three dimensions, is 115.7 Mpc. Apart from 
five member clusters, which have a redshift greater than z = 0.025 
and are labelled with squares in Figs. 1 and 2, the redshift range of the cluster members is $z = 0.0147
- 0.0223$, which corresponds to a radial extent of 32.3 Mpc (including the five clusters the maximum
extent is 71.2 Mpc). Thus, most of the Perseus-Pisces SC is oriented closely along the plane of the sky.
For an approximate volume estimate of the total SC, we 
take the sum of the spheres with radius 19 Mpc (equal to the minimum linking length) around
all member clusters accounting for the overlap, and obtain a value of $2.9 \times 10^5$ Mpc$^3$.  
Table 3 lists these properties for three different parts of the Perseus-Pisces SC
(PP = full SC, PP$_{in}$ = structure at $z \le 0.03$, PP$_{core}$ = core region with
13 members) and for the SGW.
 
Using this value for the volume, we can derive a mass estimate for the SC.
For 22 cluster members we find a cluster density of $7.6 \times 10^{-5}$ Mpc$^{-3}$
(with weights the density is $2.0 \times 10^{-4}$ Mpc$^{-3}$), which
is about a factor of 2.5 higher than the mean density of clusters with $L_X > L_{X_0}$
of $7.2 \times 10^{-5}$ Mpc$^{-3}$, determined from the mean luminosity function in a 
volume up to $z = 0.4$ \citep{Boe2014}. 
A cluster overdensity ratio of $R_{cl} = 2.8$ corresponds to a cluster overdensity 
$\Delta_{cl} = (n_{cl} - <n_{cl}>)/<n_{cl}> =1.8$. This can be related to the overdensity
of matter by means of the predicted value for the large-scale structure bias,
which depends on the lower mass limit of the cluster sample. 
From theoretical calculations \citep{Tin2010}
based on simulations and from our previous studies of clusters in the REFLEX sample, we find
a value for the cluster bias around 1.6 for the mass range involved here (for details
of these considerations see \citealt{Boe2020}). Therefore we get an overdensity for
the total matter of $\Delta_{DM} \sim 1.1$ and a matter overdensity ratio of $R_{DM} \sim 2.1$.
Together with the SC volume derived above, this leads to a 
SC mass estimate of $2.5 \times 10^{16}$ M$_{\odot}$. It can be compared
to the sum of the mass of all clusters in the Perseus-Pisces SC of $3.6 \times 10^{15}$ M$_{\odot}$,
which is only a small fraction of the SC mass. We can also compare this number to the early
estimate by \citet{Joe1978b} with a value of $2 \times 10^{16} h_{50}^{-1}$ M$_{\odot}$, which is
of a similar order of magnitude. Taking into account that our SC stretches 
into the ZoA, and is thus larger, and that we have been quite 
generous in staking out the SC volume, the two
numbers are surprisingly consistent. For smaller parts of the Perseus-Pisces SC
similar calculations can be performed and the results on the estimated SC mass are 
provided in Table 3. 

The SGW merges with the Perseus-Pisces SC if we choose a minimum
linking length of 21 instead of 19 Mpc. The shortest distance between members of each SC
is 29.8 Mpc. Thus, these are the nearest superclusters to each
other in the local Universe. 

The largest extent between two members in three dimensions in the SGW is 45.8 Mpc
and the volume for a spherical zone around each cluster with 20 Mpc radius is 
$1.2 \times 10^5$ Mpc$^3$, which yields a 
weighed cluster density of $1.2 \times 10^{-4}$ Mpc$^{-3}$,
similar to the Perseus-Pisces SC. With the same arguments as above we arrive
at a mass estimate for the SGW of $6.8 \times 10^{15}$ M$_{\odot}$. The sum of the estimated
masses of the individual clusters, with a value of  $5.0 \times 10^{14}$ M$_{\odot}$,
constitutes a fraction of $\sim 7.3\%$, which is smaller by about a factor of
two than the value for Perseus-Pisces. This is due to the fact that the SGW contains
no massive clusters. The SC properties are also summarised in Table 3.

\begin{figure}[h]
   \includegraphics[width=\columnwidth]{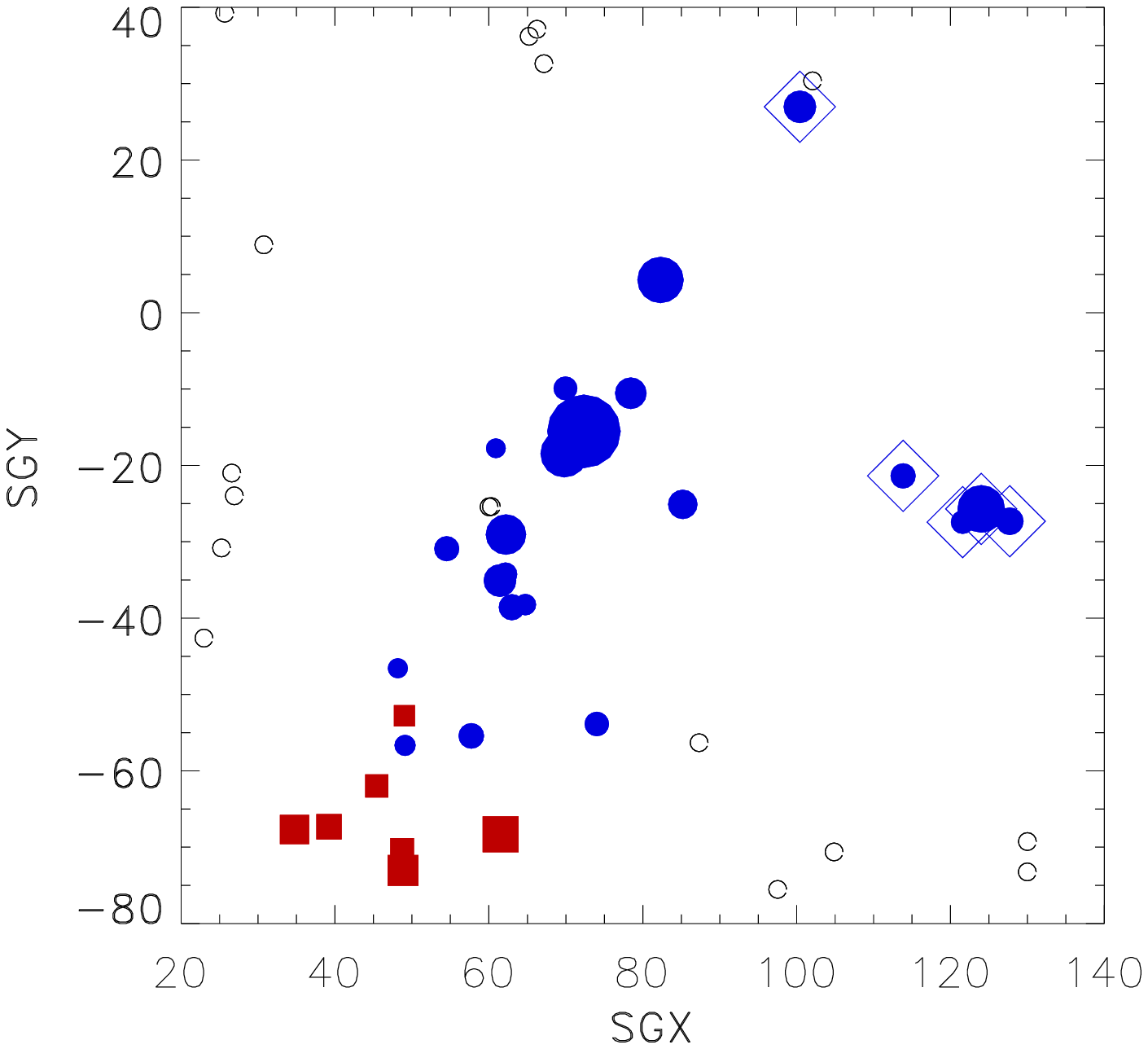}
   \includegraphics[width=\columnwidth]{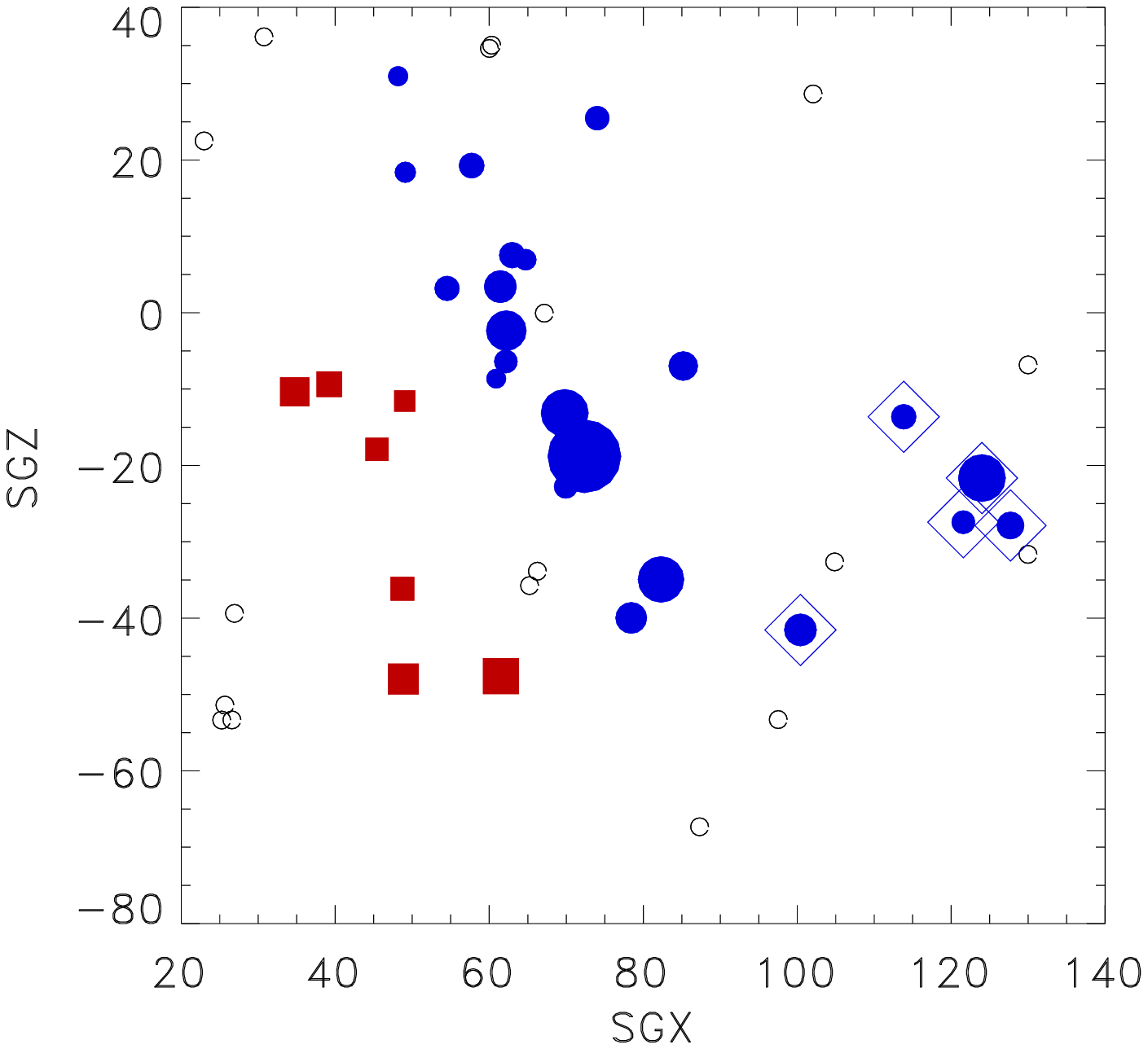}
   \includegraphics[width=\columnwidth]{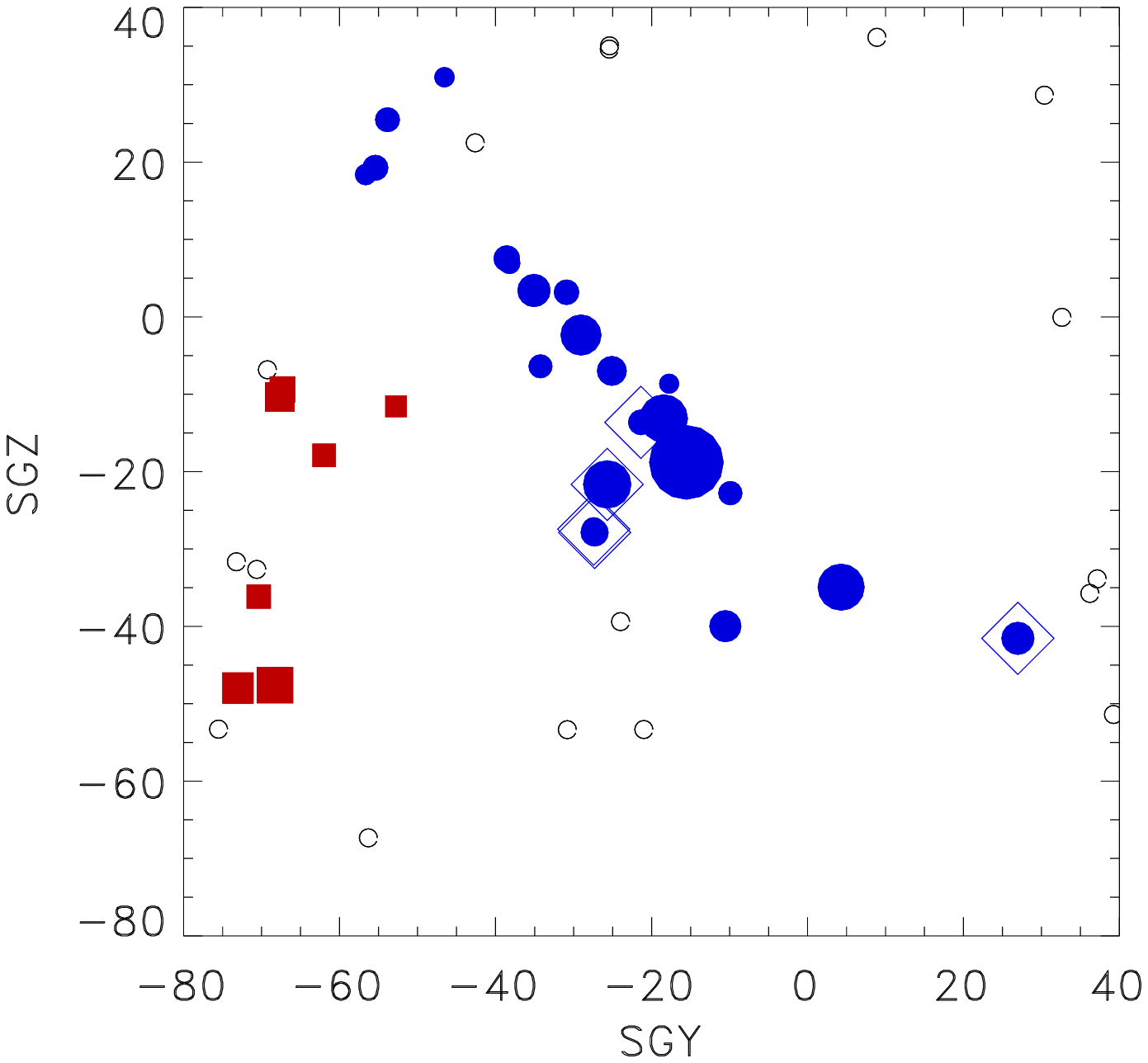}
\caption{Perseus-Pisces SC and Southern Great Wall in Supergalactic
coordinates, where the Supergalactic plane is defined by SGX and SGY. 
The members of the Perseus-Pisces SC are shown by blue filled circles, 
and those of the Southern Great Wall by red squares.  
}\label{fig3}
\end{figure}

\begin{figure}[h]
   \includegraphics[width=\columnwidth]{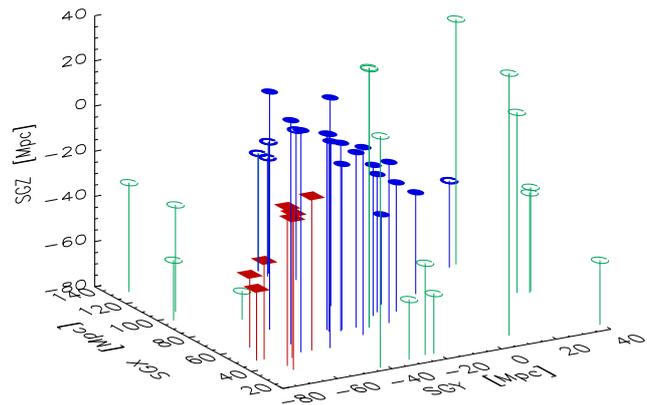}
\caption{Three-dimensional representation of the {\sf CLASSIX} cluster 
distribution in a box of 120 Mpc side length around the Perseus-Pisces SC
in Supergalactic coordinates as indicated on the axes of the plot. The members
of the Perseus-Pisces SC are shown by filled blue circles and those of
the Southern Great Wall by filled red diamonds, while all other clusters are 
marked by open circles. 
}\label{fig6}
\end{figure}

\subsection{Alternative linking schemes}

In a subsequent paper of this series we will explore the SC
construction  for various lower X-ray luminosity limits. While keeping the goal of the selection
of structures with an overdensity ratio of at least 2 fixed, we studied how the
selection of the SCs change when restricting the cluster sample to objects
with higher minimum X-ray luminosity. In general, the overall structure selection is quite
robust, but in detail the linking of some extensions of the structures can change.

When we choose a higher X-ray luminosity limit for the cluster sample in this approach,
we need to adjust the linking length to the more sparse cluster density if we want to keep the 
minimum overdensity ratio constant. Thus, on the one hand, there is a possibility to  link to new
X-ray luminous clusters. On the other hand, connections relying on low luminosity
objects can disappear. If, for the construction of the Perseus-Pisces SC, we make an extreme 
change by increasing the lower X-ray luminosity limit by a factor of 10 to
$10^{43}$ erg s$^{-1}$, we get the change illustrated in the lower panel of Fig.~\ref{fig1}.  
In the figure the groups and clusters linked by the default construction 
to form the Perseus-Pisces SC are shown
by solid blue dots. The clusters selected at the high luminosity limit are marked
by large blue, open circles. We note that the main chain of the Perseus-Pisces SC
is traced well in both cases. However, while at low $L_{X_0}$ there is  an extension 
consisting of a small group of low luminosity objects in the south-west, two clusters
towards the north-west are linked at high $L_{X_0}$. The latter extension is 
interesting in comparison to the galaxy distribution, as discussed in section 5. 

\subsection{Three dimensional structure}

As a representation of the three-dimensional structure of the Perseus-Pisces SC
and the Southern Great Wall, we show their structure in Supergalactic Cartesian
coordinates in three projections in Fig.~\ref{fig3}. The members of the 
Perseus-Pisces SC are shown as blue filled circles, and those of the SGW
as red filled squares. The five clusters of the Perseus-Pisces SC with 
$z > 0.025$ are marked by additional embedding open diamonds. We note that
they form a clearly separated group of systems, as shown by two of the projections.

With respect to the Supergalactic plane, the major axis of the Perseus-Pisces SC
has an inclination of about 34$^o$. Within the Supergalactic plane the SC
main axis has an angle of about 36$^o$ with the X-axis towards the positive Y-axis. The
Perseus cluster occupies a  central position in the SC. The SGW
has a lager extent perpendicular to the Supergalactic plane than within
the plane.

In Fig.~\ref{fig6} we show a three-dimensional representation of the
two SC. We note that the two SCs are close but clearly separated.
The distribution of groups and clusters in this figure can be compared
to the three-dimensional matter distribution inferred from peculiar motions
of galaxies in the local Universe as studied by the group of Tully et al. 
(e.g. \citep{Cou2013,Tul2014,Tul2019}). A good illustration for the 
comparison is Fig. 8 by \citet{Cou2013}, where we see clearly the mass 
concentrations of the Perseus-Pisces SC and the Southern Great Wall
in a very similar configuration to that  shown in Fig.~\ref{fig6}.

Among the nearby SC the Perseus-Pisces SC stands out with its remarkable morphology.
It features a long chain of groups and clusters in an almost straight and 
linear structure. On the sky the main axis of this SC has a position angle of
$\sim 107^o$ (for details of the determination of the angle see Appendix A3).

Earlier studies of the Perseus-Pisces SC, for example the one by \citet{Joe1978b},
have found that the shapes of the clusters are aligned with the elongation
of the SC. We can test if this is also supported by X-ray observations by
studying the shapes of the X-ray images of the SC members. This is explained in
more detail in Appendices A1 and A2. In the Appendices we study the alignment in two 
respects. First, we study the alignment of the cluster shapes with the main axis 
of the SC, and then we inspect whether the clusters are aligned with the
direction of the tidal pull of all other cluster members.
In general, we see  no evidence of the two types of alignment. 

A very clear alignment is found, however, in the shapes of the Perseus cluster
and the neighbouring group AWM7. The ellipticity of both systems is well aligned
with the separation vector of the two. The alignment, which is almost
horizontal (position angle $\sim 90^o$), is only weakly aligned with the SC
with position angle 107$^o$. 

\subsection{Member clusters}

\begin{figure}[h]
   \includegraphics[width=\columnwidth]{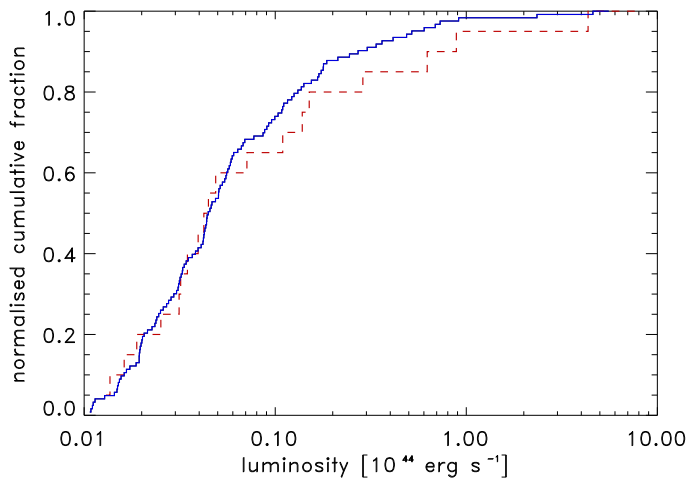}
\caption{Cumulative normalised distribution of the X-ray luminosity
of the groups and clusters of galaxies in the Perseus-Pisces SC (at $z \le 0.03$)
(red dashed line) compared to the luminosity distribution of the
systems outside the Perseus-Pisces SC and the Southern Great Wall (blue solid line).
}\label{fig7}
\end{figure}

In our previous studies of superstes-clusters, which have a larger 
overdensity, we found that the mass or X-ray luminosity distribution 
of the member clusters is more top heavy than the 
distribution of these properties of clusters in the field.
Would we find the same effect in these larger structures at lower overdensity? 
Figure~\ref{fig7} shows the cumulative, normalised X-ray luminosity distribution of 
the groups and clusters in the Perseus-Pisces SC compared to that of clusters in
the surrounding volume at $z \le 0.03$ (also not contained in the Southern Great Wall).
We see an interesting trend, that the Perseus-Pisces SC contains more luminous clusters.
A Kolmogorov-Smirnov test shows, however, that this result  
(77\% probability that the two distributions agree) based on just 
one SC is not significant.

\section{Discussion}

We took a new approach to characterise the Perseus-Pisces SC and the Southern Great Wall
with X-ray luminous clusters. In addition to finding a very similar structure of the SC
to that found in previous optical studies,
we now see that it extends through the ZoA. This makes the Perseus-Pisces SC the largest 
SC in the local Universe at $z \le 0.03$, with more groups and cluster 
members than even the Southern Great Wall. It also makes the largest contribution to the 
flattened local superstructure extending the Local Supercluster to a radius of about 
100 Mpc, which we studied previously \citep{Boe2021}.

Based on radio observations of neutral hydrogen, the galaxy distribution in the 
surroundings of the Perseus-Pisces SC was studied, extending the redshift survey
into the ZoA. \citet{Hau1987} and \citet{Cha1990} investigated an extension of the 
Perseus-Pisces SC across the ZoA, testing an earlier suggestion by \citet{Bur1979}
of an extension up to the cluster A569 (see Fig.~\ref{fig1}, bottom). 
Both redshift surveys find a set of galaxies
around 3C129 connecting it to the Perseus-Pisces SC. While \citet{Hau1987}
argues that the SC seems to end there, \citet{Cha1990} find some more redshifts indicating
 a connection to A569. \citet{Ram2016} find a larger number of galaxies in a halo around
3C129 at the same distance as the Perseus-Pisces SC, further strengthening this connection.
As shown above, we find in our study that the Perseus-Pisces SC extends through the cluster,
3C 129 close to the galactic equator, to the other side of the ZoA to UGC 3355. A569 
has 3C129 as its closest neighbour with a distance of 36.2 Mpc, and  thus too far 
away for the given linking length.

An additional cluster was discussed in these previous works as
a member of the  Perseus-Pisces SC located at the west side of the SC is A2534 (see Fig.~\ref{fig1}, 
bottom). This cluster is even further away from the nearest SC member with a distance of 
41.8 Mpc. While it appears in Fig.~\ref{fig1} near to a small western group of
SC members, its higher redshift of $z = 0.0312$ is responsible for the distance and
it is not linked to the SC, even if we relax the condition for members to be at $z \le 0.03$.

Another interesting point is a comparison to the galaxy distribution characterising
the Perseus-Pisces SC including many redshifts in the ZoA from the work by
\citet{Kra2018}. They show (in their Fig. 10) a very prominent 
filament of galaxies roughly following the chain of galaxy groups and clusters 
from UGC 3355 in the east to UGC 12655 in the west. This galaxy filament
then continues by turning north and crossing the ZoA roughly
at galactic longitude $90^o$. Even though this extension of the galaxy filament is
also prominent, it is not traced by a chain of groups and clusters.
However, approximately in the middle of this extension we find  a group of X-ray 
luminous systems, which indicate at least  one location in this filament. Two of these
clusters are luminous enough that they are linked to the Perseus-Pisces SC
with a higher low luminosity cut and corresponding to a larger linking length, as explained
in section 4.3. This seems to be the only indication marking the western upturn of the
SC as traced by clusters.

Characterising the shapes of the groups and clusters in the Perseus-Pisces SC as 
traced by the X-ray emission, we find no significant alignment of cluster elongations
with the main axis of the SC, except for the effect that Perseus and AWM7 are 
pointing to each other. This is different from earlier findings using optical surveys.
A major reason for this could be that the X-ray emission is focussed on the central, virialised,
and thermalised part of the clusters, which by nature has acquired a  rounder shape. This
effect is further strengthened as the X-ray emission follows roughly equipotential contours
that are   rounder than the galaxy and dark matter distribution. In the study by 
\citet{Buo1996} one can typically find, for example,   ellipticity values that are  three times higher 
 in the projected dark matter distribution compared to the X-ray surface brightness.
Therefore, even though X-rays are a very sensitive probe of the cluster structure, they
are not particularly able to capture the large-scale ellipticity of a cluster.
It would be interesting in this context to study the shape of the member clusters
of the Perseus-Pisces SC in the optical separating the tracer galaxies into a red and blue or early-
and late-type population. In our previous studies (e.g. \citealt{Sch1999}, for the
Virgo cluster, and \citealt{Bra2009}), we find that the red early-type galaxy 
population follows the X-ray emission very closely, whereas the blue galaxies
 preferentially populate the outskirts. Therefore, we would expect that
alignment effects with the SC would most strongly show up in the blue galaxy population.

Another difference between our work and studies based on optical surveys of the galaxy 
population is the inclusion of compact groups with one dominant giant elliptical
(some of them qualify as fossil groups), which would be very inconspicuous in
the optical. However, this choice is closest to choosing dark matter halos with a given
mass limit with the mass measured inside a certain overdensity radius, which in
our case is $r_{500}$. One of the most compact of the groups in our sample is NGC 410.
As an example we give more details on this object in Appendix A3. The X-ray emission
of the group can be resolved with XMM-Newton. Fitting a $\beta$-model, we find
a core radius of the X-ray image
of $\sim 5$ arcsec. The group was also studied by \citet{Osu2017}
who confirm that the X-ray emission of the group is from a thermal intracluster
plasma with a temperature of $\sim 0.98$ keV. We can trace the X-ray emission
in the XMM-Newton observation out to about 150 arcsec ($\sim 54$ kpc), whereas
the estimate for $r_{200}$ is about 419 kpc. Thus, for some of these very compact 
groups we observe only the inner core of the system. This is also one more reason why it
is difficult to detect the large-scale structure alignment for these low mass
systems in X-rays.

On the other hand, two groups and clusters have been associated with the Perseus-Pisces
SC by \citet{Joe1978b}, A347 and NGC315, which are not X-ray luminous enough
to be included in our study. A347 was also assigned to the Perseus-Pisces SC
by \citet{Gre1981}. In these two objects we observe no X-ray emission in
the RASS for A347, for which we can set an upper limit on the X-ray luminosity
of $L_X < 0.9 \times 10^{42}$ erg s$^{-1}$, and for NGC315 we find a faint X-ray 
source with an upper limit of $L_X < 0.8 \times 10^{42}$ erg s$^{-1}$.
Since X-ray observations probe the relaxed core of clusters, the non- or
faint detections of the two poor clusters imply masses for their
relaxed parts of $m_{200} \le 2 \times 10^{13}$ M$_{\odot}$.

\section{Summary and conclusion}

One of the major goals of this paper is to provide a detailed reference for
the structure and properties of the Perseus-Pisces SC and the neighbouring
Southern Great Wall as traced by X-ray luminous clusters. We find that
the Perseus-Pisces SC with a length of $\sim 116$ Mpc and a mass of the order
of $2.5 \times 10^{16}$ M$_{\odot}$ is the largest SC in the local 
Universe out to a redshift of $\sim 0.03$. In our study we assign 22 groups
and clusters with a total mass of about $3.6 \times 10^{15}$ M$_{\odot}$ to this
SC (including 2 clusters at $z = 0.03 - 0.0314$).  The SGW is about two and a half
times smaller in length and three times smaller in mass and  number of 
members. 

The two SCs  defined in this work
have a cluster density that is larger by a 
factor of $\sim 2.5$ compared to the mean density in the study volume,
which implies a matter overdensity ratio
of about 2. Thus, both structures are well within the non-linear regime,
which also explains their highly elongated shapes. However, their overdensity is
not nearly high enough for these structures to collapse to one unit
in the future in a $\Lambda$CDM universe.
Instead, they will  fragment and form
a series of very massive clusters in the future.

The Perseus-Pisces SC is a remarkable straight chain 
of galaxy groups and clusters,
which extends even further in one direction than previously 
inferred from optical surveys, because  we can now trace it 
through the ZoA. Previous neutral hydrogen surveys had already
given indications of this.
The SC is dominated by the Perseus cluster, while the Southern 
Great Wall is not inhabited by massive clusters. 

The shapes of the X-ray images of the Perseus-Pisces SC members are not found 
to align with the major axis of the SC. The Perseus cluster and AWM7
are, however, well aligned with each other.

\begin{acknowledgements}  We thank the referee for helpful comments.
HB and GC acknowledge support of the Deutsche 
Forschungsgemeinschaft through the Munich Excellence Cluster ``Universe''. 
G.C. acknowledges support by the DLR under grant no. 50 OR 1905.
\end{acknowledgements}

\appendix
\section{Online material}

\subsection{X-ray - Optical images of the Perseus-Pisces supercluster members}

Figures A.1 - A.4 provide images of the member groups and clusters of the Perseus-Pisces
SC. They consist of X-ray surface brightness contours superposed on optical
images from the DSS digitisation of the photographic sky surveys. 
The X-ray contours are based on XMM-Newton observations when available. 
In this case all three detectors
were combined with a scaling of the pn-images with a factor of 3.3 with respect
to the MOS images. For systems with no XMM-Newton observations,
we use X-ray images from the ROSAT All-Sky Survey. An exception is the Perseus 
cluster, where the field of view of XMM-Newton is too small to show the entire cluster
in one pointing. For this reason, we use a ROSAT pointed observation for the Perseus 
cluster.

\begin{figure*}[h]
\hbox{
\hspace{1cm}
   \includegraphics[width=7.2cm]{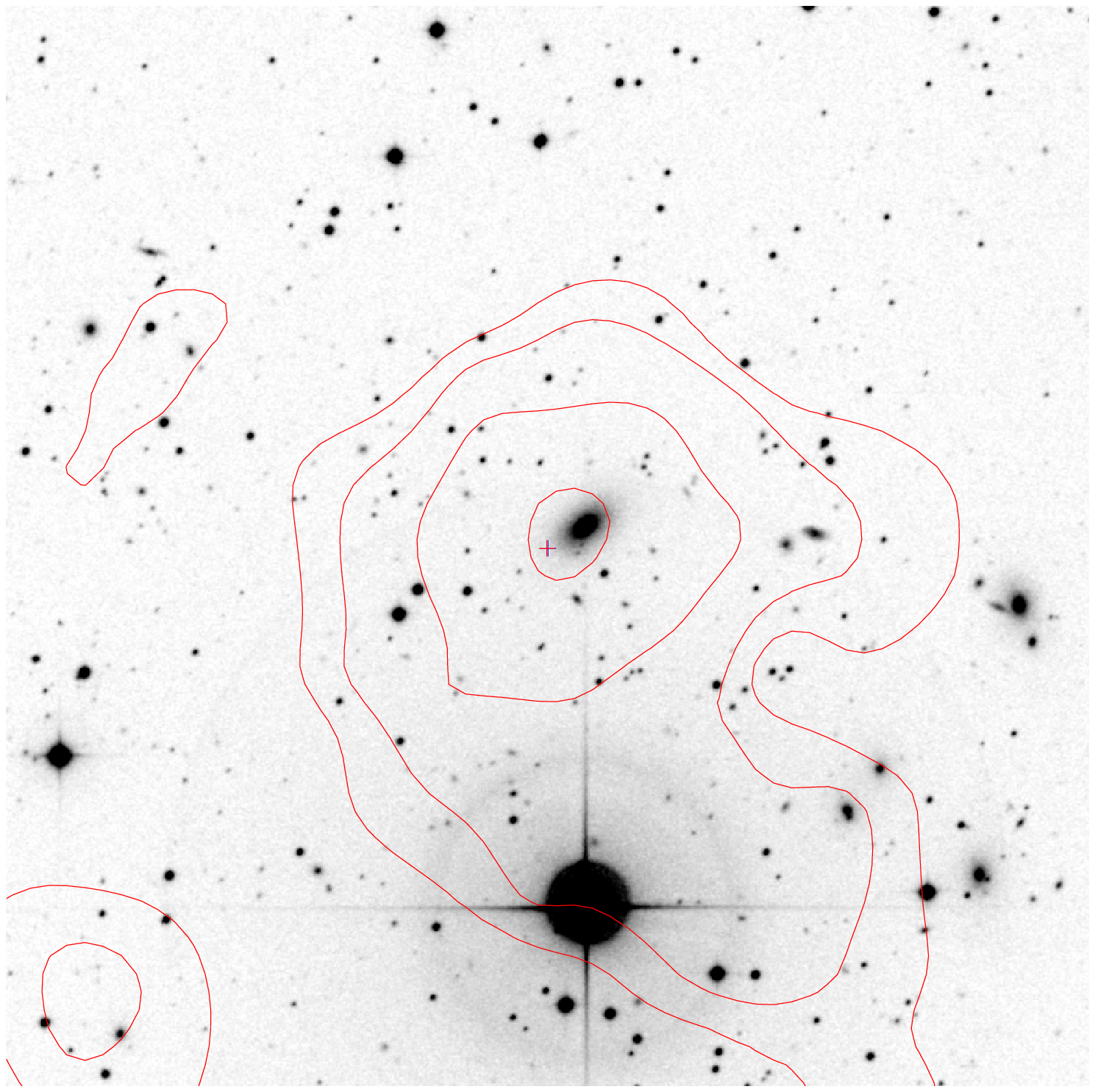}
\hspace{1cm}
   \includegraphics[width=7.2cm]{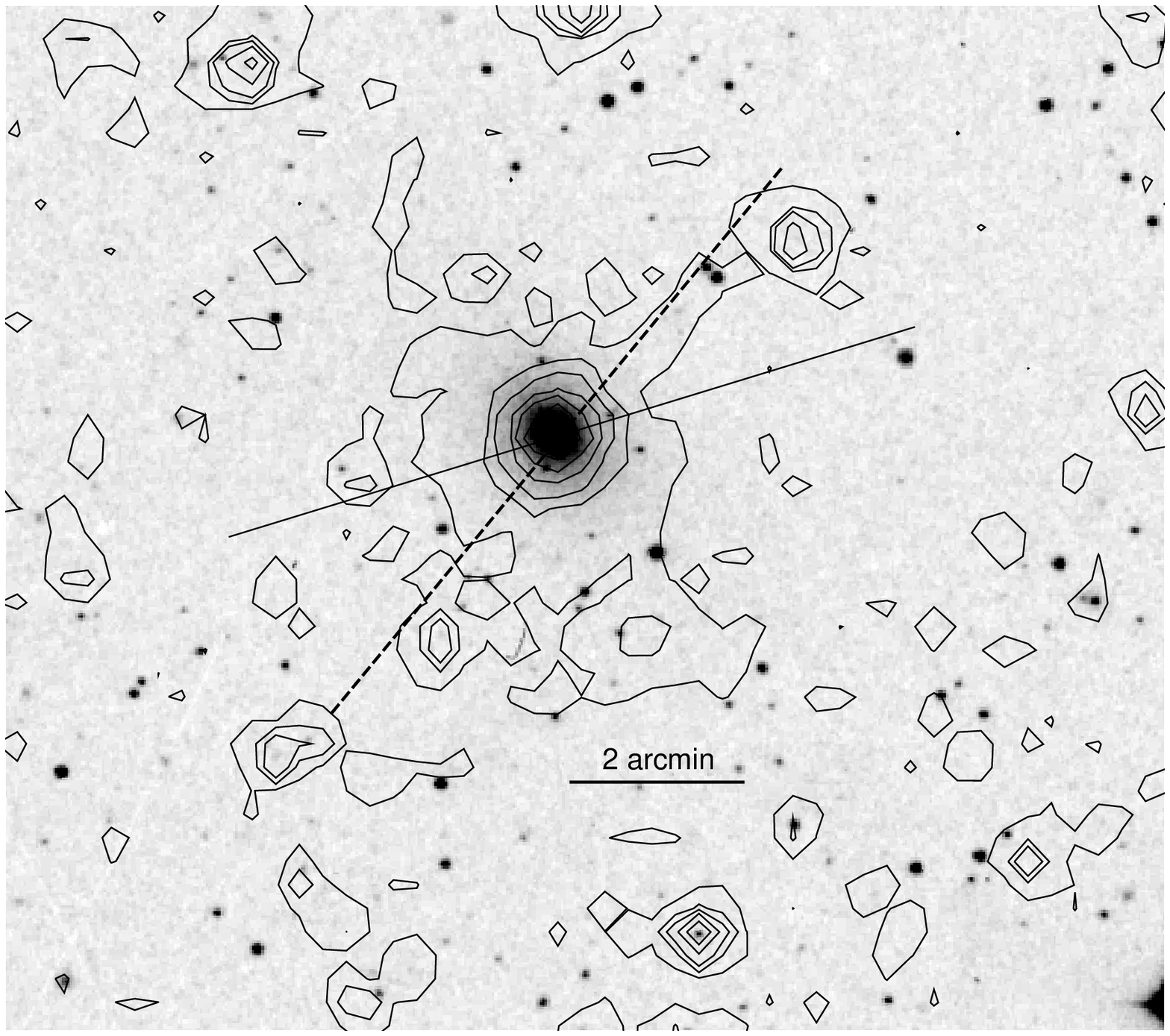}
}
\hbox{
\hspace{1cm}
   \includegraphics[width=7.2cm]{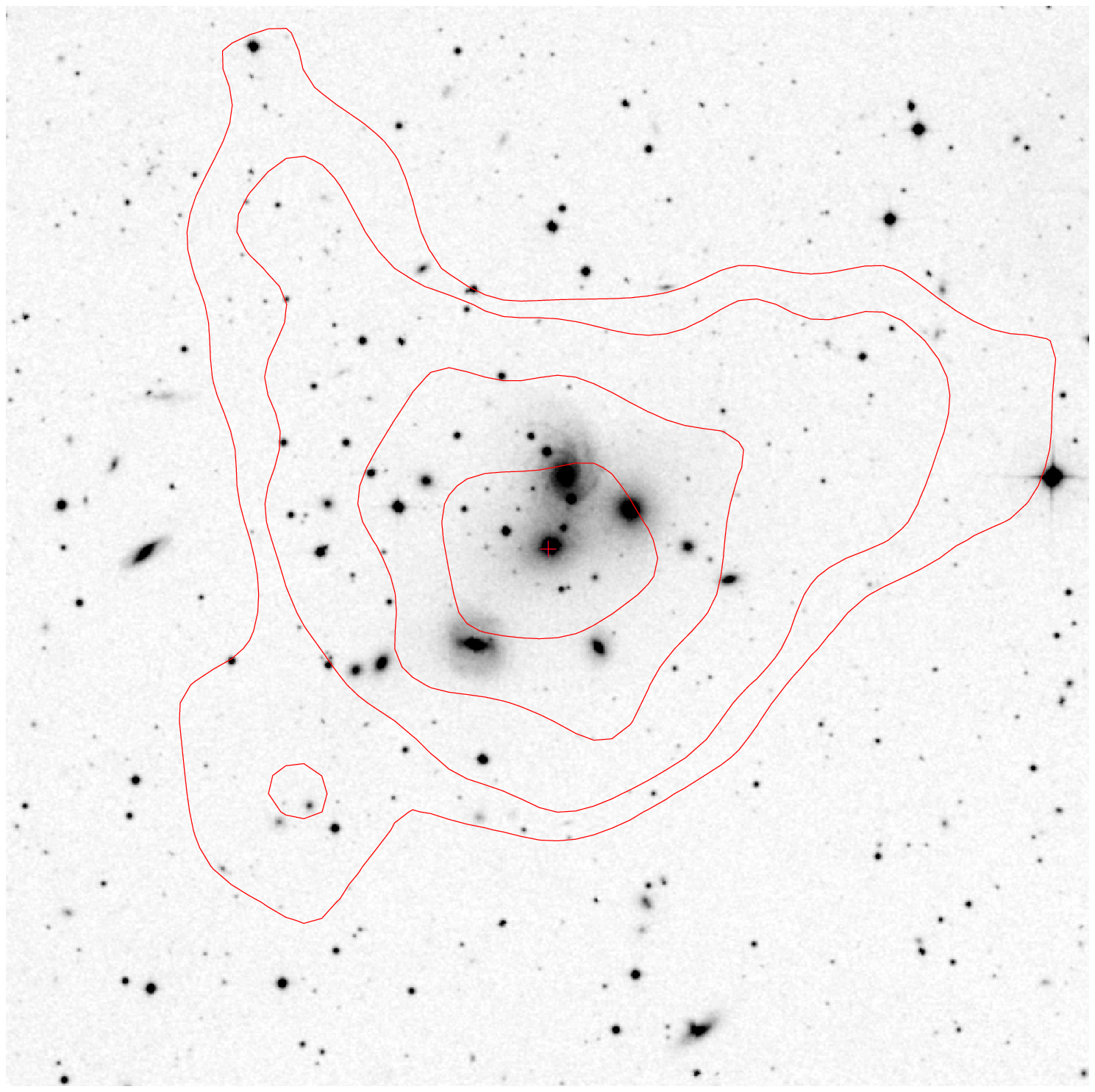}
\hspace{1cm}
   \includegraphics[width=7.2cm]{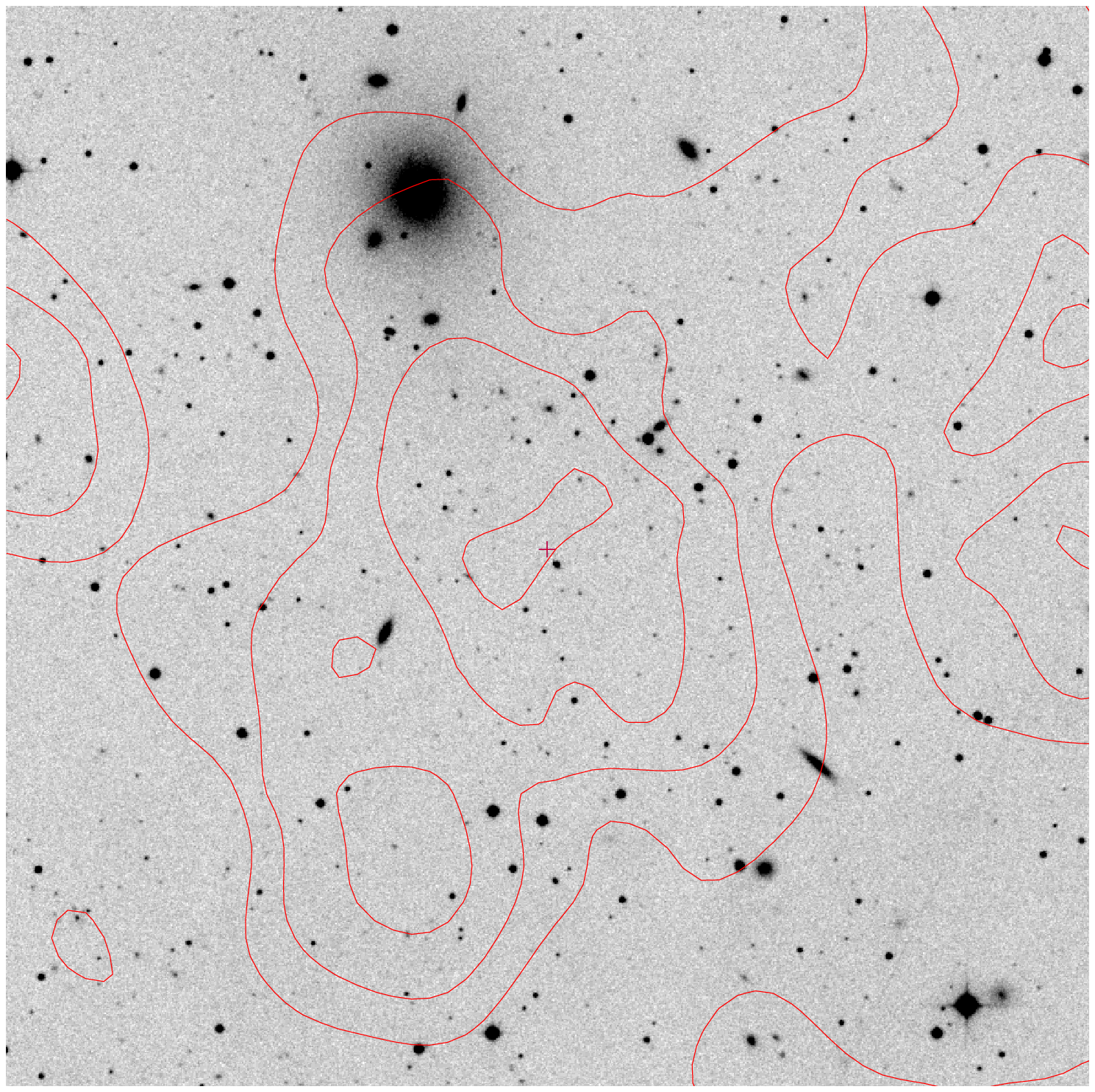}
}
\hbox{
\hspace{1cm}
   \includegraphics[width=7.2cm]{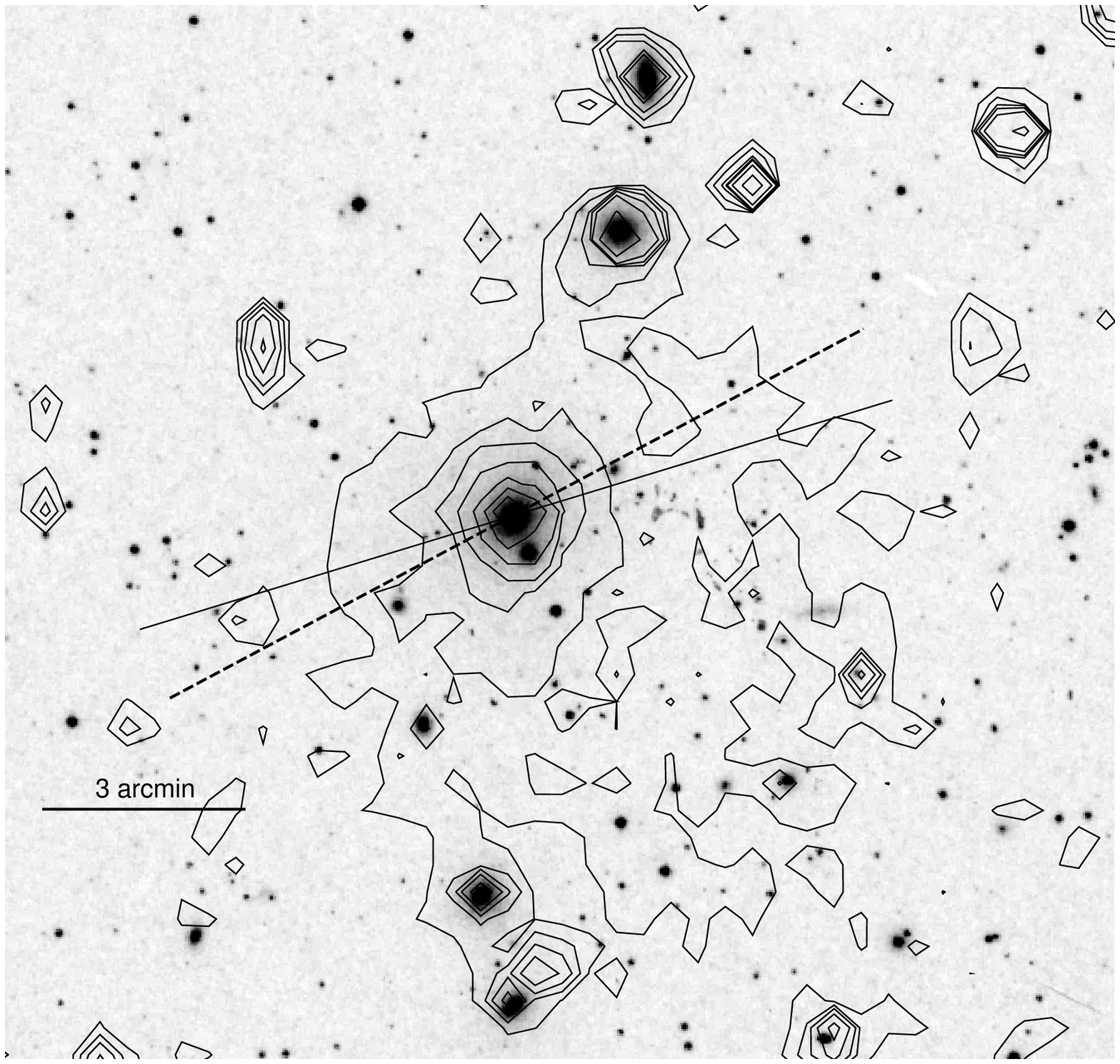}
\hspace{1cm}
   \includegraphics[width=7.2cm]{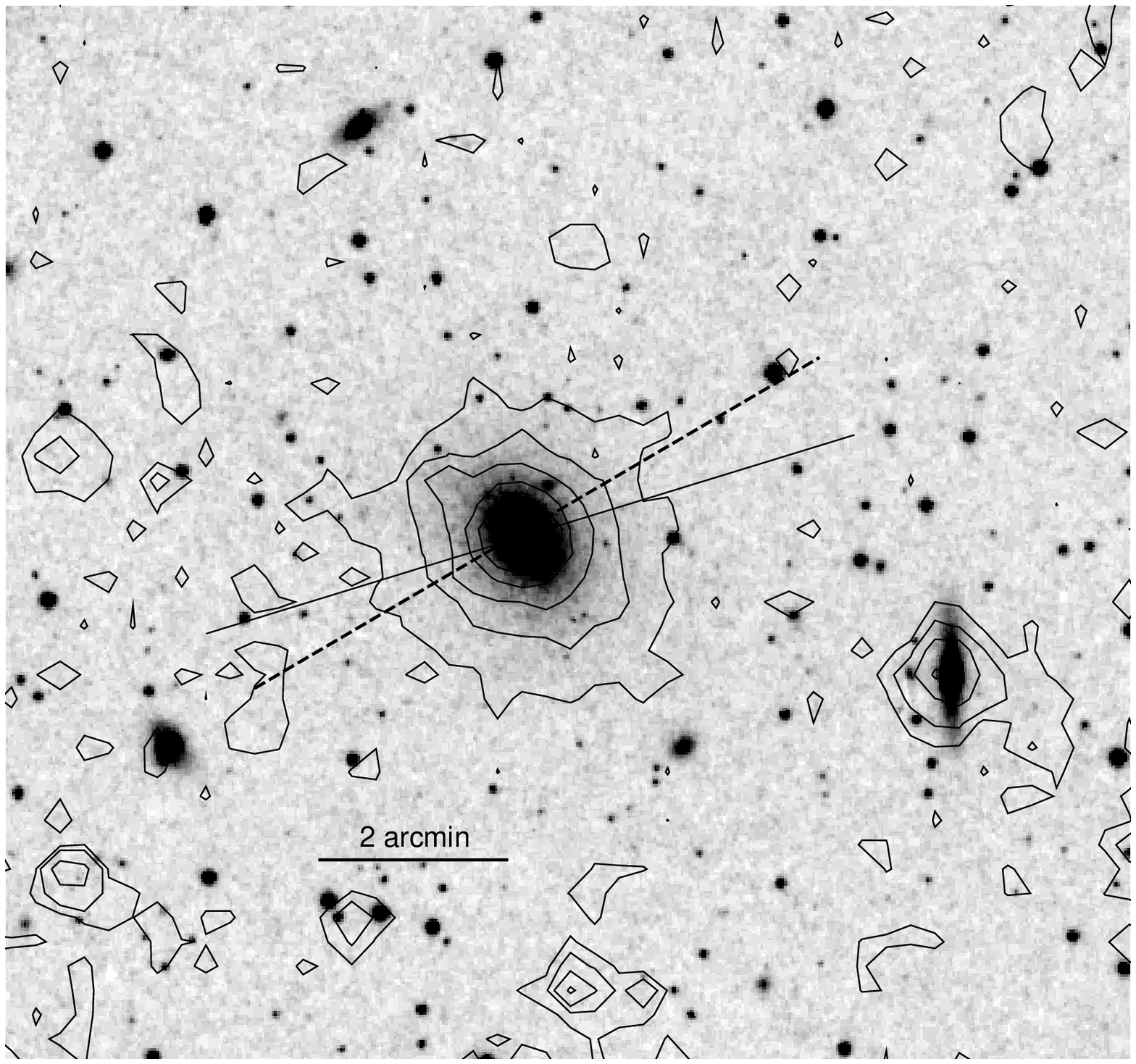}
}
\caption{Contours of the X-ray surface brightness superposed on optical images 
from the DSS database. The X-ray data are  from the RASS or XMM-Newton observations.
All images here and in the following figures are oriented such that north is 
up and and east to the left.
{\bf Upper left:} RXCJ2332.5+2355, UGC12655, X-ray data from RASS; 
{\bf Upper right:} RXCJ0015.5+1720, NGC 57, X-ray data from XMM-Newton; 
{\bf Middle left:} RXCJ0018.3+3003, NGC 71, X-ray data from RASS; 
{\bf Middle right:} RXCJ0021.0+2216, SRGb063, PPS 62, X-ray data from RASS; 
{\bf Lower left:} RXCJ0107.2+3224, NGC 383; X-ray data from XMM-Newton; 
{\bf Lower right:} RXCJ0110.9+3308, NGC 410, X-ray data from XMM-Newton.
}\label{figA1}
\end{figure*}

Overplotted we show the direction of the main axis of the
Perseus-Pisces SC and the direction of the combined tidal forces of all other SC
members on the clusters for comparison with the cluster shapes.
For the RASS images we do not show these directions, as the photon statistics
are too poor for a reliable determination of the cluster or group elongation. 
Appendix A2 provides details of how these directions were determined.

\begin{figure*}[h]
\hbox{
\hspace{1cm}
   \includegraphics[width=8cm]{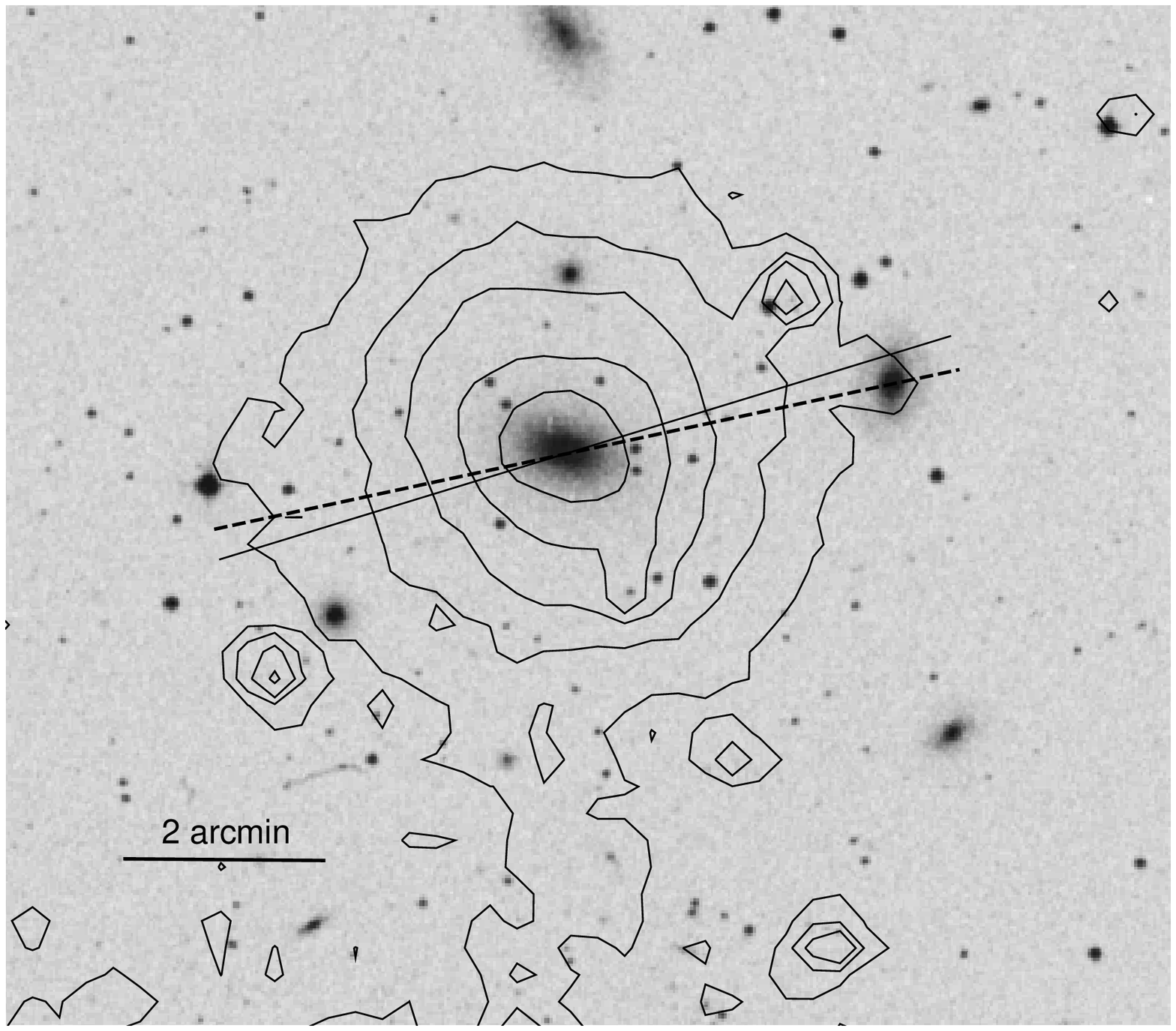}
\hspace{1cm}
   \includegraphics[width=8cm]{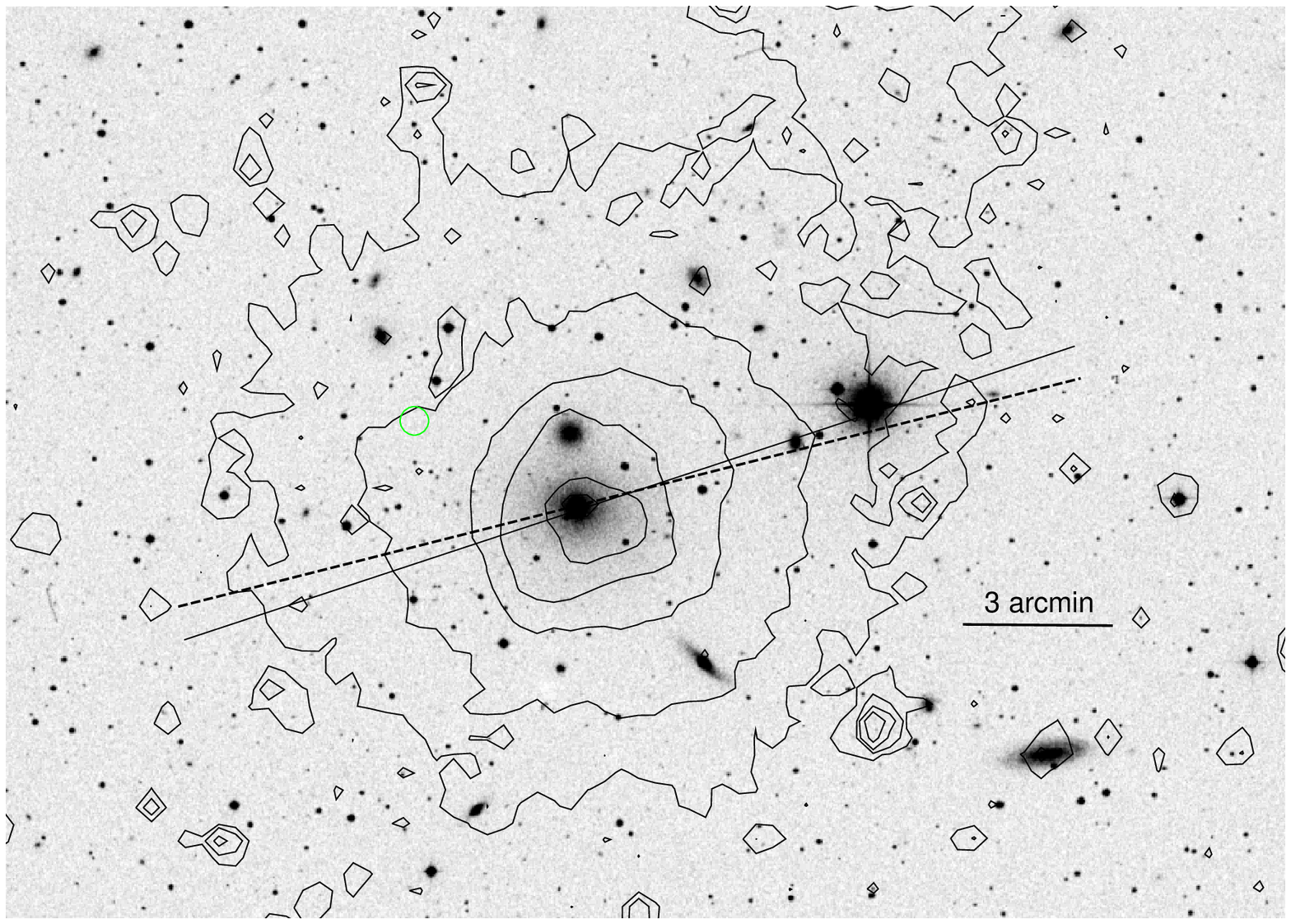}
}
\hbox{
\hspace{1cm}
   \includegraphics[width=8cm]{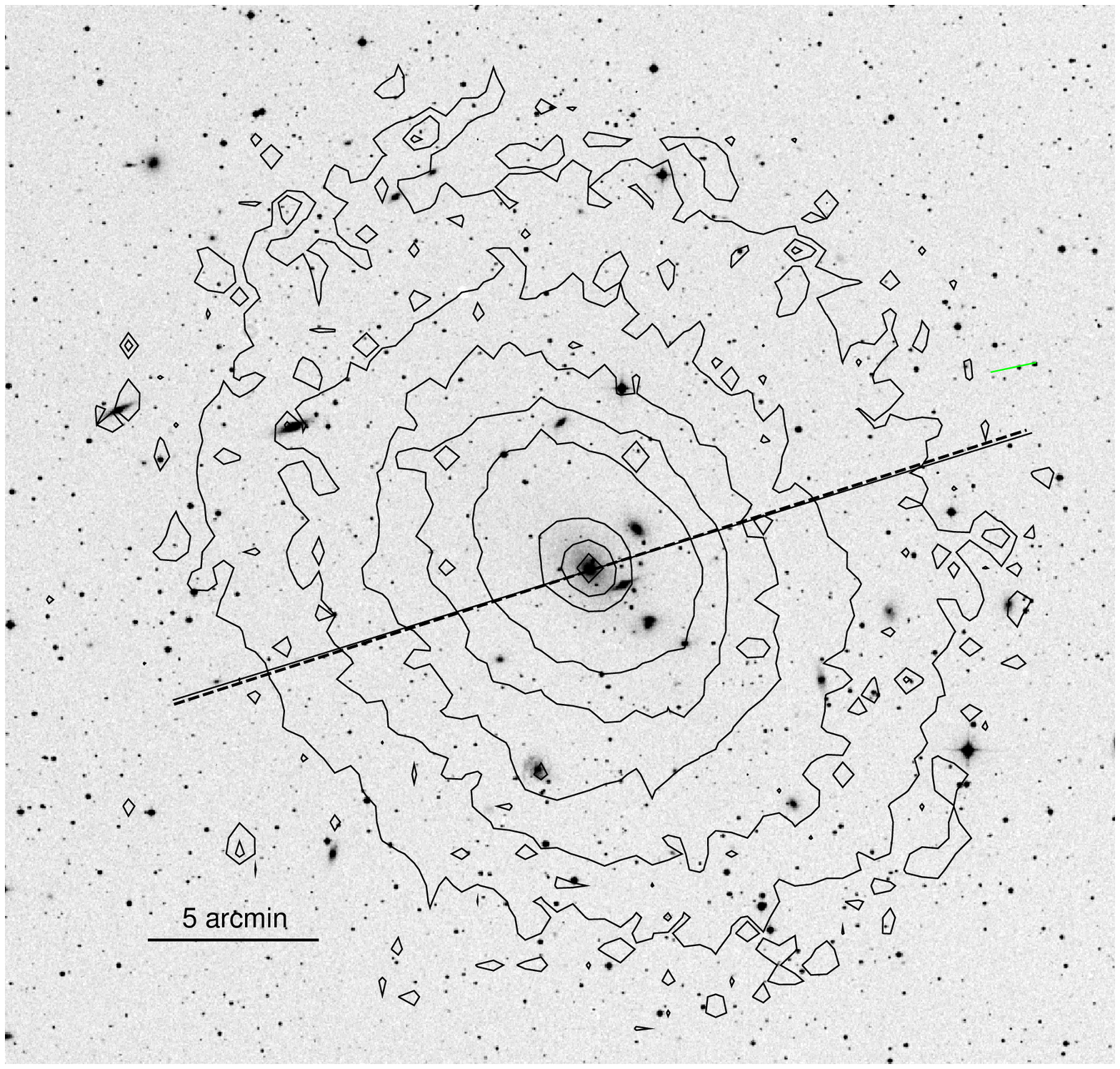}
\hspace{1cm}
   \includegraphics[width=8cm]{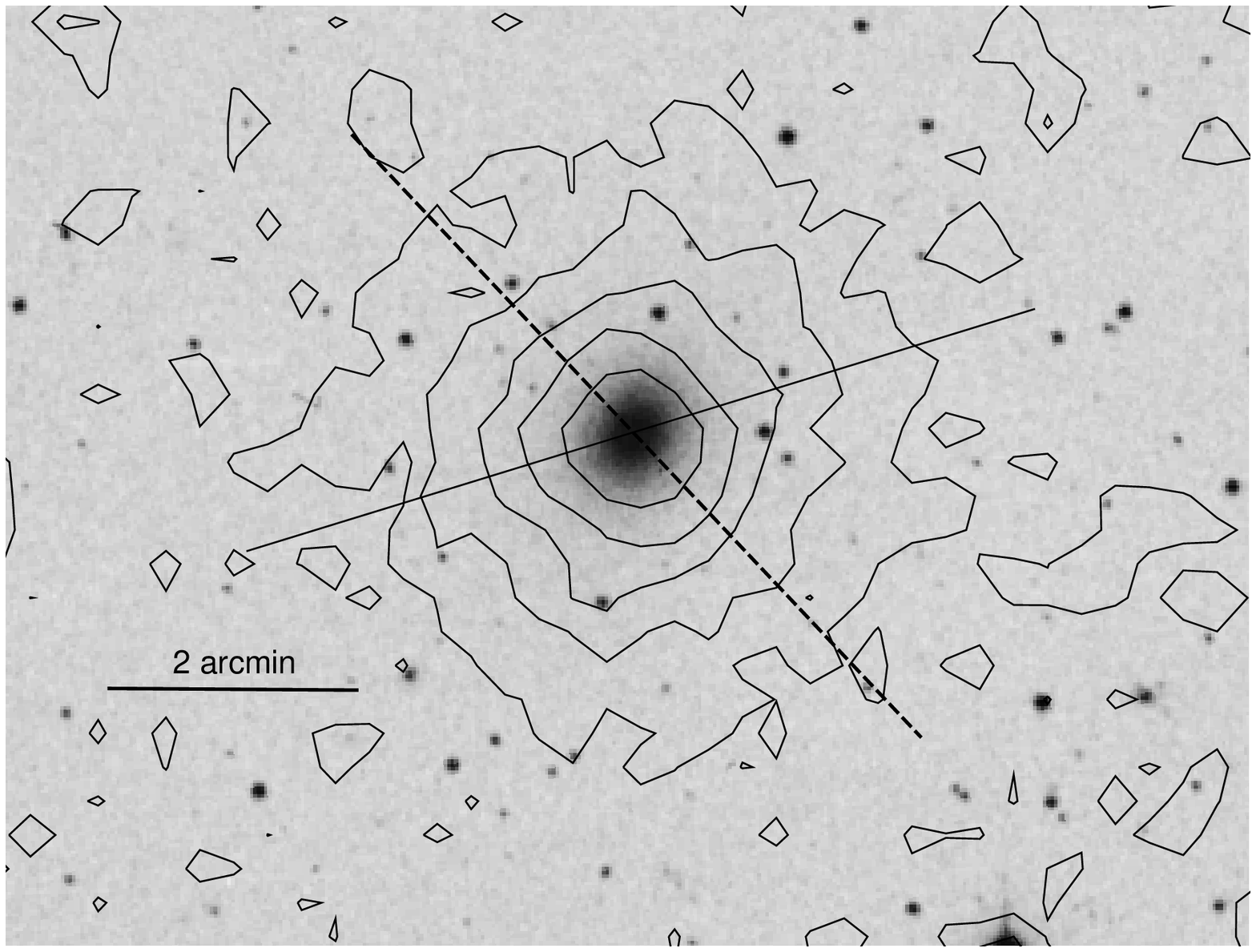}
}
\hbox{
\hspace{1cm}
   \includegraphics[width=8cm]{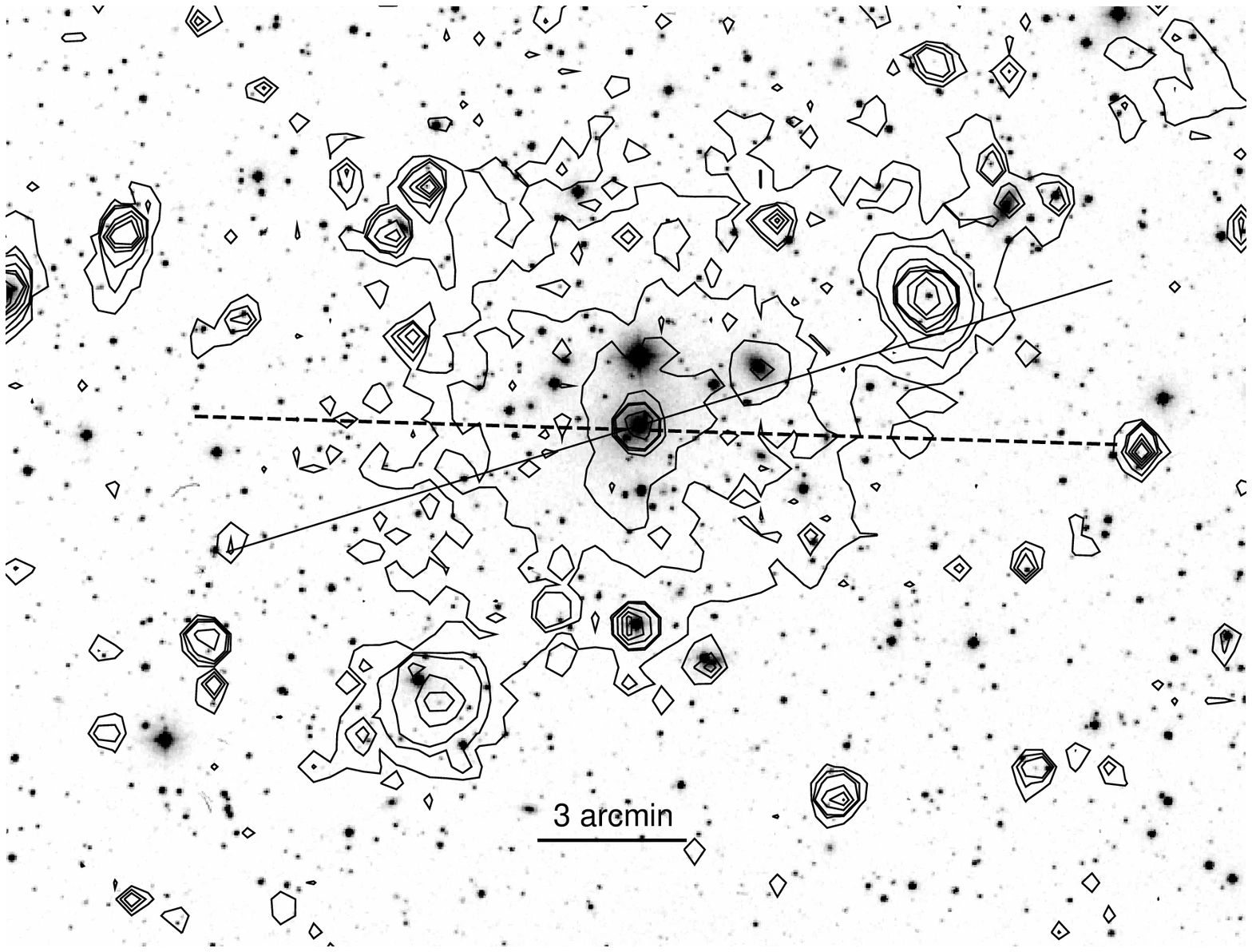}
\hspace{1cm}
   \includegraphics[width=8cm]{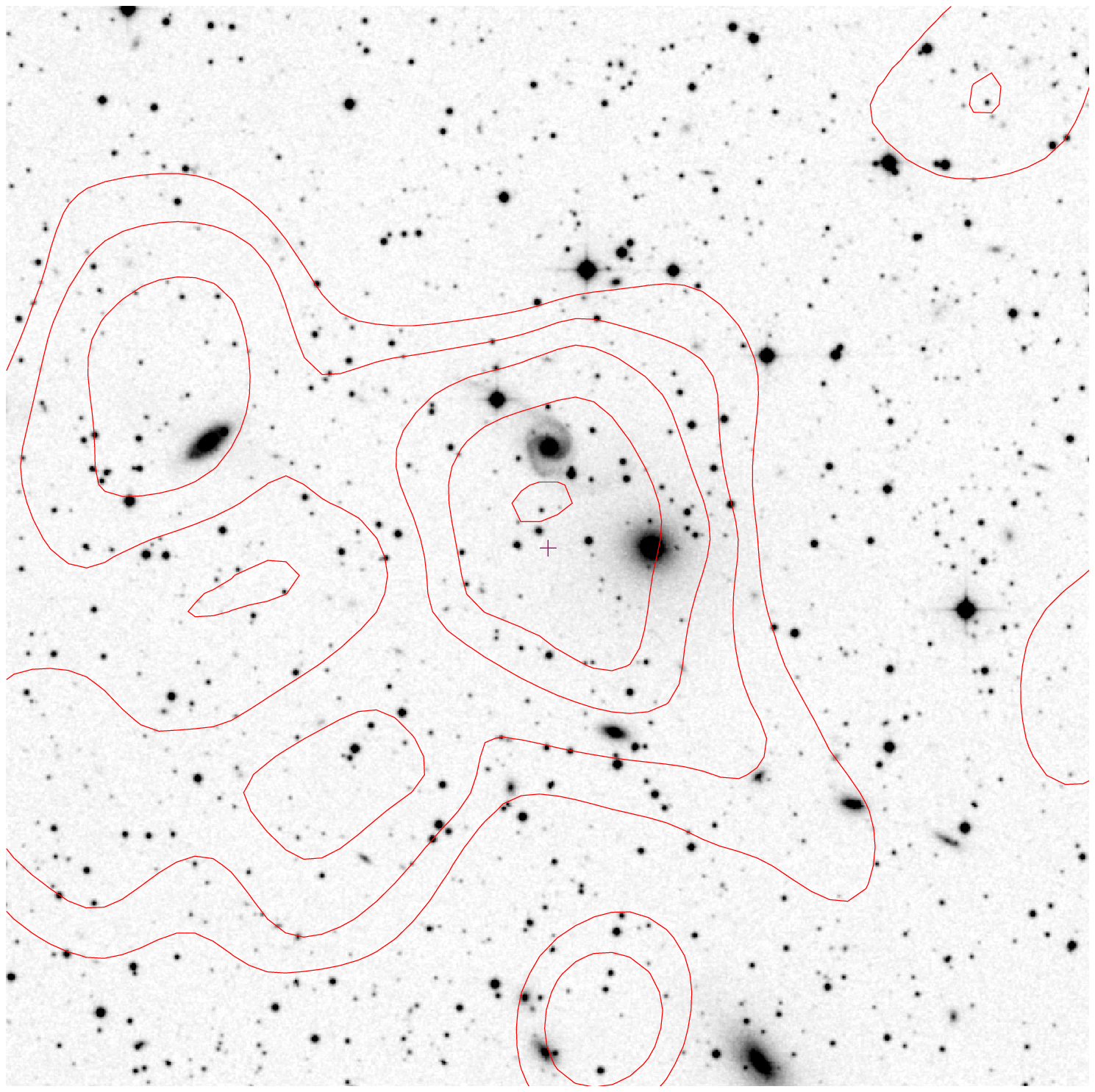}
}
\caption{Images of Perseus-Pisces SC members continued.
{\bf Upper left:} RXCJ0123.1+3327, NGC 499, X-ray data from XMM-Newton; 
{\bf Upper right:} RXCJ0123.6+3315, NGC 507, X-ray data from XMM-Newton; 
{\bf Middle left:} RXCJ0152.7+3609, A 262, X-ray data from XMM-Newton; 
{\bf Middle right:} RXCJ0200.2+3126, NGC 777, X-ray data from XMM-Newton; 
{\bf Lower left: } R0222.7+4301, UGC 1841, X-ray data from XMM-Newton;
{\bf Lower right:} RXCJ0238.7+4138, NGC 996, X-ray data from RASS.
}\label{figA1}
\end{figure*}

\subsection{Alignment of galaxy clusters with the supercluster}

To investigate whether the cluster shapes in the Perseus-Pisces SC are aligned
with the cluster elongation, we determined the main axis of the SC by deriving
the tensor of the moment of inertia and its principle axis in projection onto the sky.
If we determine the major axis of the inertia tensor with just the object
distribution or with the objects weighed by their mass, we obtain 
in both cases a position angle for the major axis of about $PA \sim 107^o$. 

To see whether the possible orientation of the cluster shapes follows the tidal
forces of the SC members, we also determined the moment of inertia and
its major axis for each cluster position scaling the distance to each other 
cluster in a special way. For the conventional moment of inertia the 
mass points introduce a weight that is proportional to the square of the
distance, while the gravitational force depends on the inverse of the distance squared.
Therefore, we scale every SC member with its mass and with an additional factor
of the distance to the inverse fourth power, $\Delta^{-4}$, in the tensor 
of the moment of inertia.

The results of both calculations are illustrated in Fig. A.5, where we show 
the major axis of the Perseus-Pisces SC as a whole by the dotted line
and the direction of the tidal forces for each SC member by short
solid lines. Both directions are displayed in the images of those
Perseus-Pisces SC members in Figs. A.1 to A.4, which have
deeper pointed observations  by XMM-Newton or ROSAT. As discussed
in the main text, we see no significant evidence of an alignment
of the cluster shapes with the SC. Both Perseus and AWM7 experience 
mutual tidal forces without much disturbance by the other SC members,
as can be seen in Fig. A.5. Here we see that the cluster shapes 
follow the direction of the tidal forces very closely.

\begin{figure*}[h]
\hbox{
\hspace{1cm}
   \includegraphics[width=7.7cm]{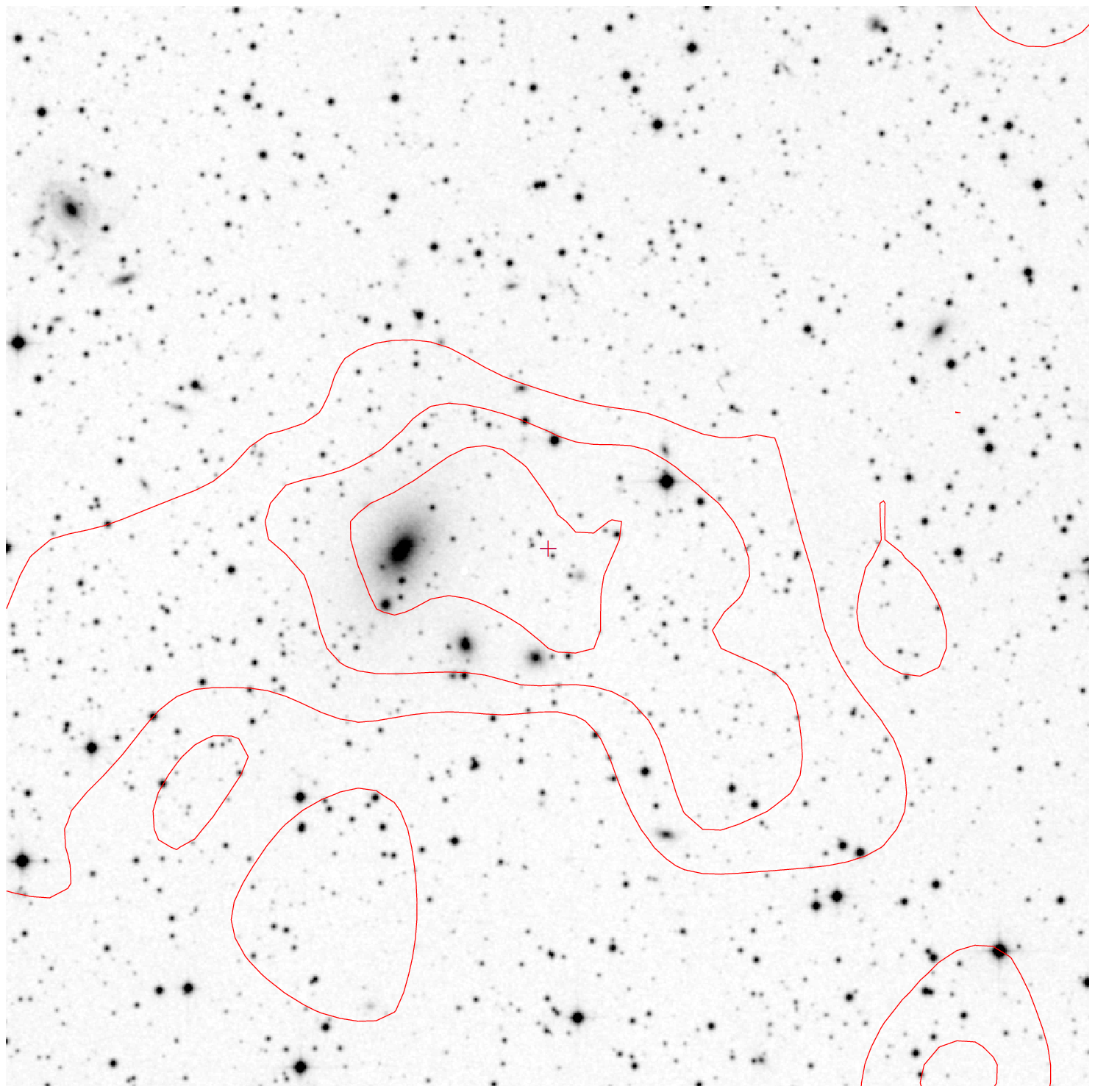}
\hspace{1cm}
   \includegraphics[width=7.7cm]{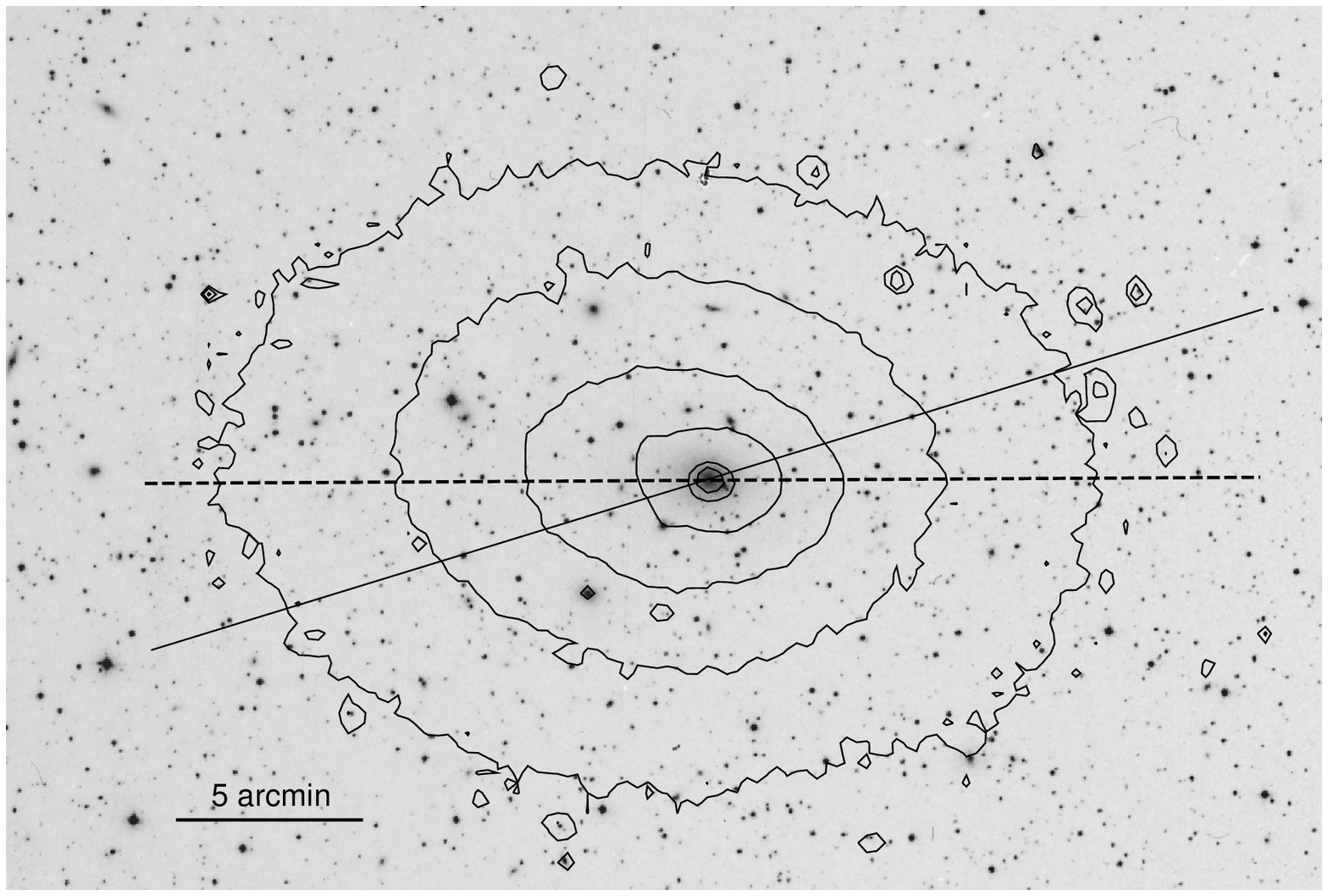}
}
\hbox{
\hspace{1cm}
   \includegraphics[width=7.7cm]{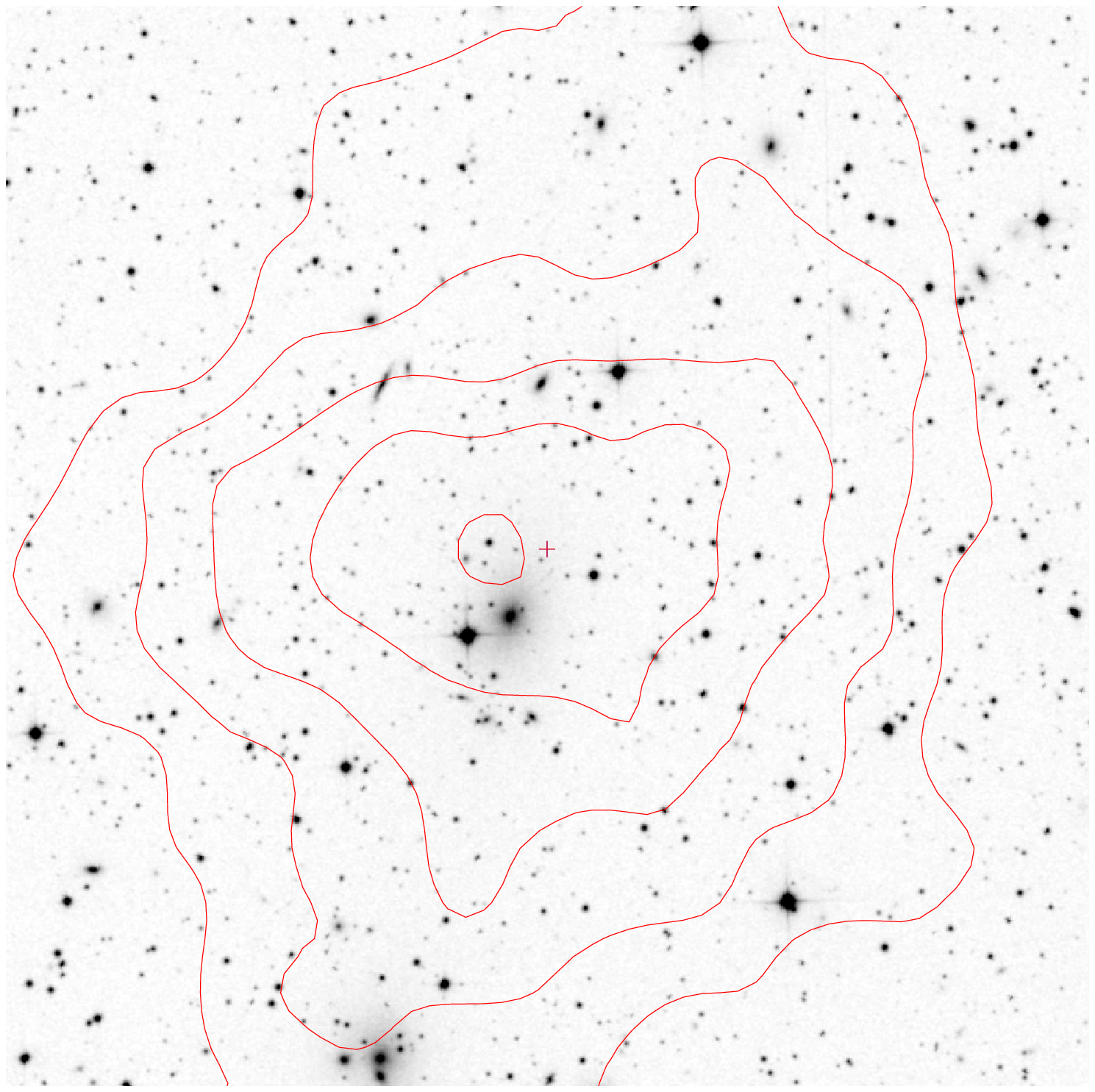}
\hspace{1cm}
   \includegraphics[width=7.7cm]{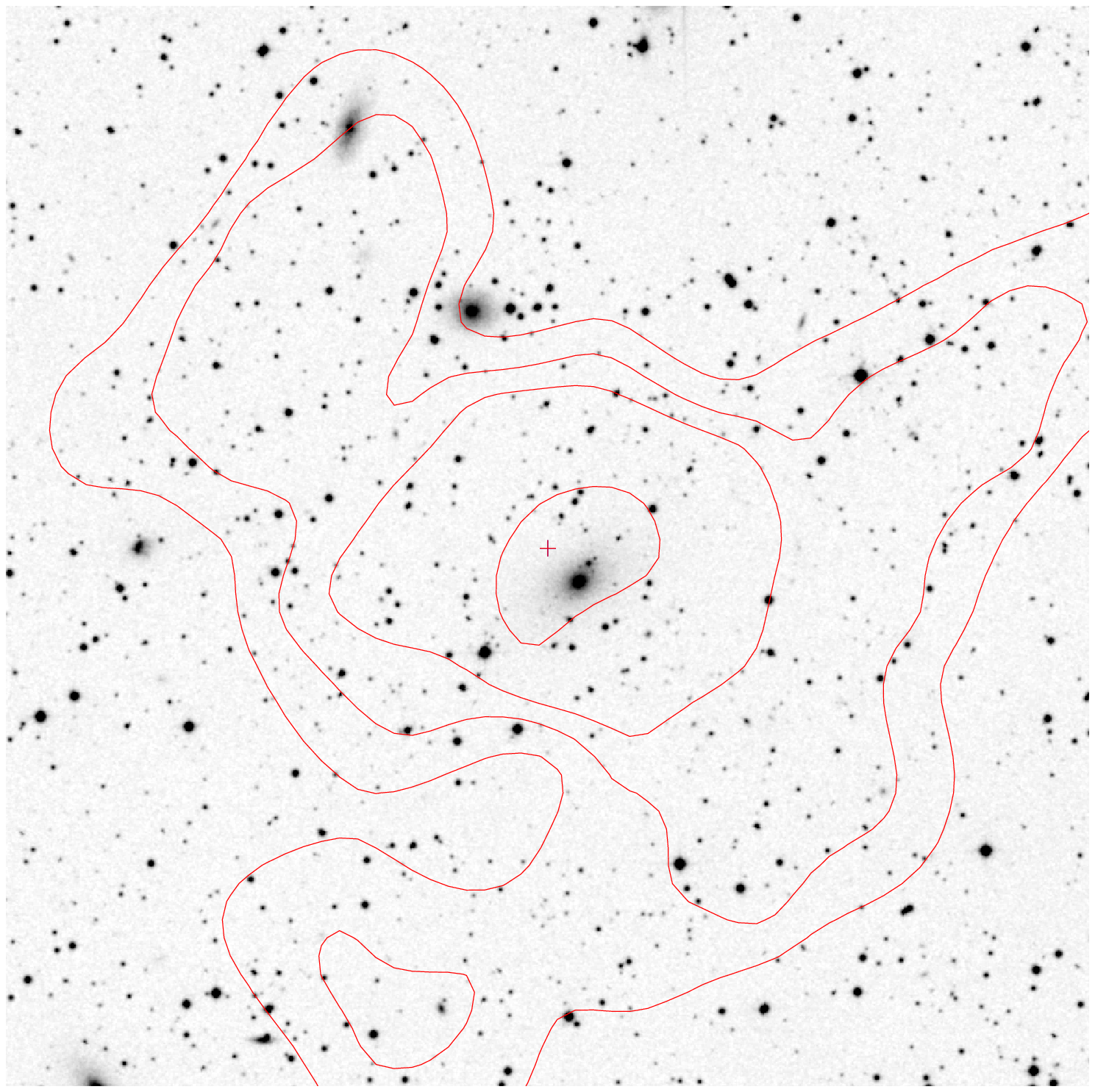}
}
\hbox{
\hspace{1cm}
   \includegraphics[width=7.7cm]{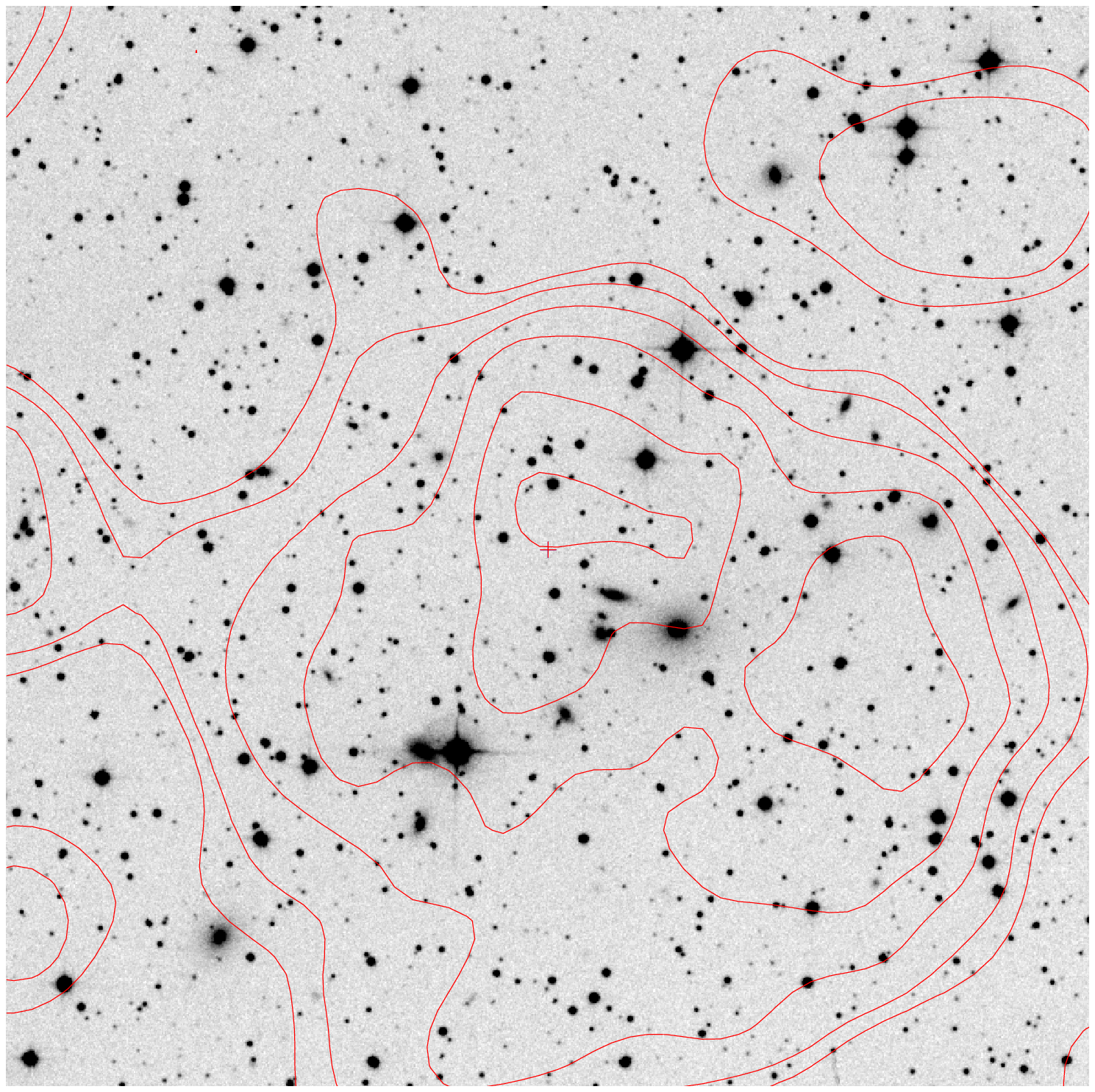}
\hspace{1cm}
   \includegraphics[width=7.7cm]{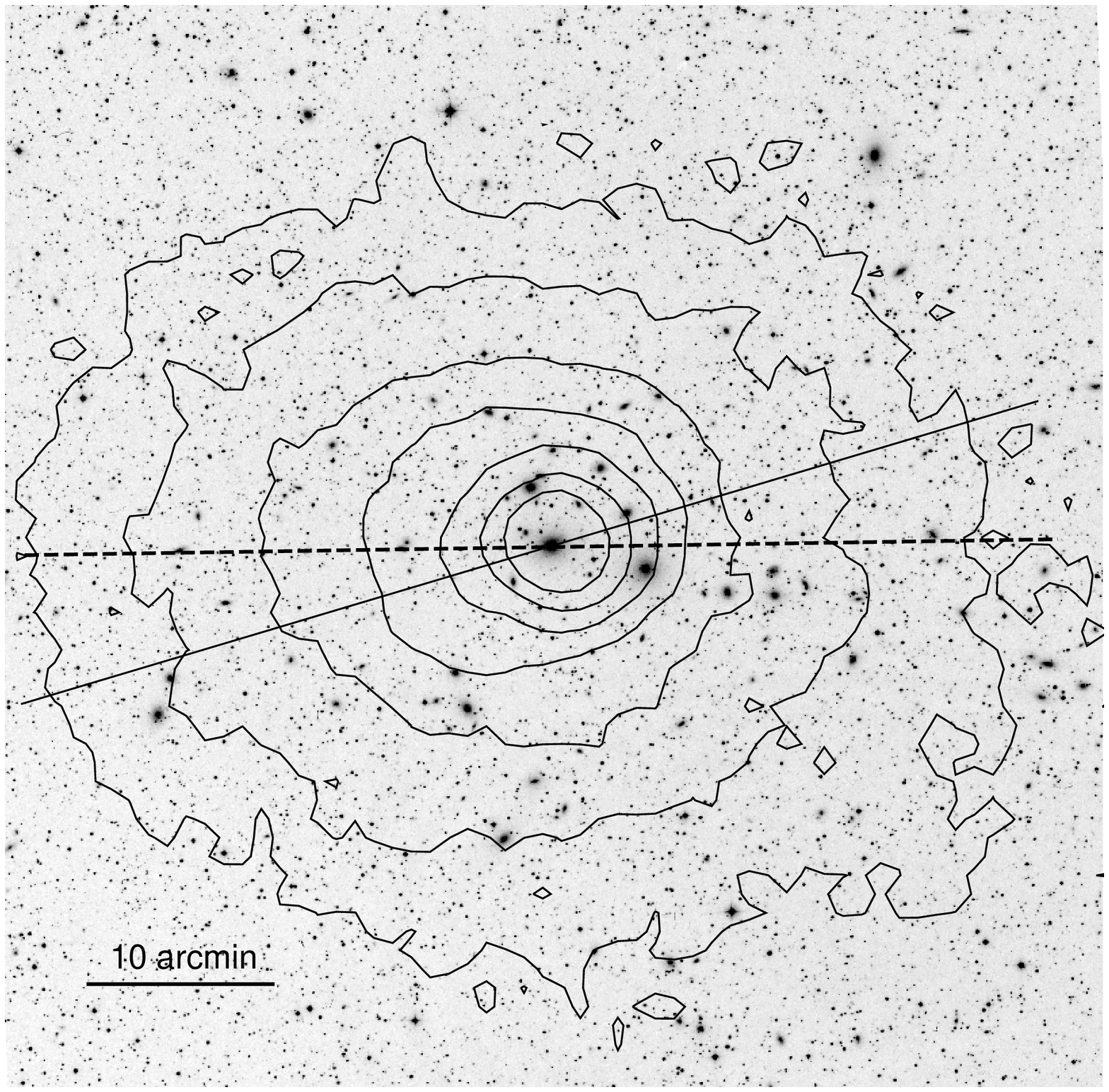}
}
\caption{Images of Perseus-Pisces SC members continued.
{\bf Upper left:} RXCJ0249.5+4658, IC 257, X-ray data from RASS;
{\bf Upper right:} RXCJ0254.4+4134, NGC 1129, AWM7, X-ray data from XMM-Newton;
{\bf Middle left:} RXCJ0300.7+4428, CIZAJ0300.7+4427, X-ray data from RASS;  
{\bf Middle right:} RXCJ0309.9+4207, UGC 2562, X-ray data from RASS;   
{\bf Lower left: } RXCJ0310.3+4250, X-ray data from RASS; 
{\bf Lower right:} RXCJ0319.7+4130,  A 426, Perseus, X-ray data from 
ROSAT pointed observation. 
}\label{figA1}
\end{figure*}

\subsection{Identification of RXCJ0021.0+2216}

One of the groups associated with the Perseus-Pisces SC, RXCJ0021.0+2216,
has a different appearance (Fig.~A.1, middle right) than typically found for the 
other systems. The X-ray emission is extended and has a diffuse,
low surface brightness. However,  the X-ray emission is not peaked on a dominant
elliptical galaxy, in this case  NGC 80 at $z = 0.01947$, which is the
bright galaxy seen in the north of the image. Therefore, it is worth  looking at the
identification of this source in more detail. In a detailed survey of X-ray 
emission of galaxy groups in the CfA redshift survey by \citet{Mah2000},
this system was identified as the X-ray luminous group SRGb063 at $z = 0.0189$.  
The NASA Extragalactic Database (NED) lists nine galaxies with redshifts 
$z = 0.019 \pm 0.0013$ within a region of 7 arcmin radius around the centre 
of the emission, well inside the extent of the X-ray emission, consistent
with this result. The group was also identified with the galaxy
group found in the optical in the Perseus-Pisces Group Survey
by \citet{Tra1998} with the name PPS 62 and a redshift
of $z = 0.0188$. In the same region \citet{Wen2009} found a galaxy
cluster at $z=0.1991$, which is supported by only one redshift in NED.

The extended shallow X-ray emission is more consistent with a low redshift group
than with a more distant X-ray luminous galaxy cluster, which should appear
more compact. The spectral X-ray hardness ratio is consistent with thermal
emission from intragroup or intracluster medium. Based on the combined
evidence, we therefore identify the X-ray source  with the low luminosity
group, which makes it a member of the Perseus SC.

\begin{figure*}[h]
\hbox{
\hspace{1cm}
   \includegraphics[width=8cm]{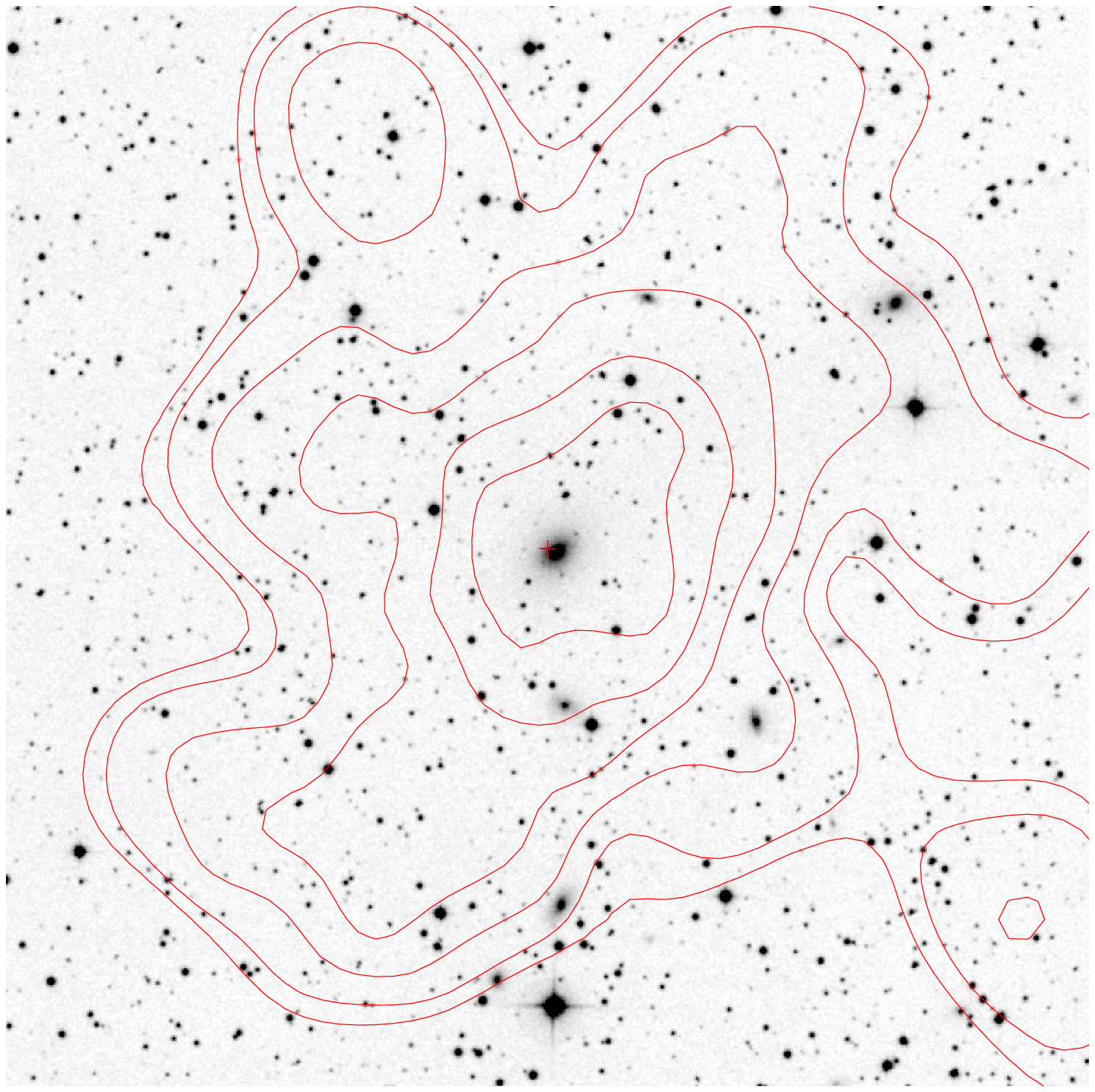}
\hspace{1cm}
   \includegraphics[width=8cm]{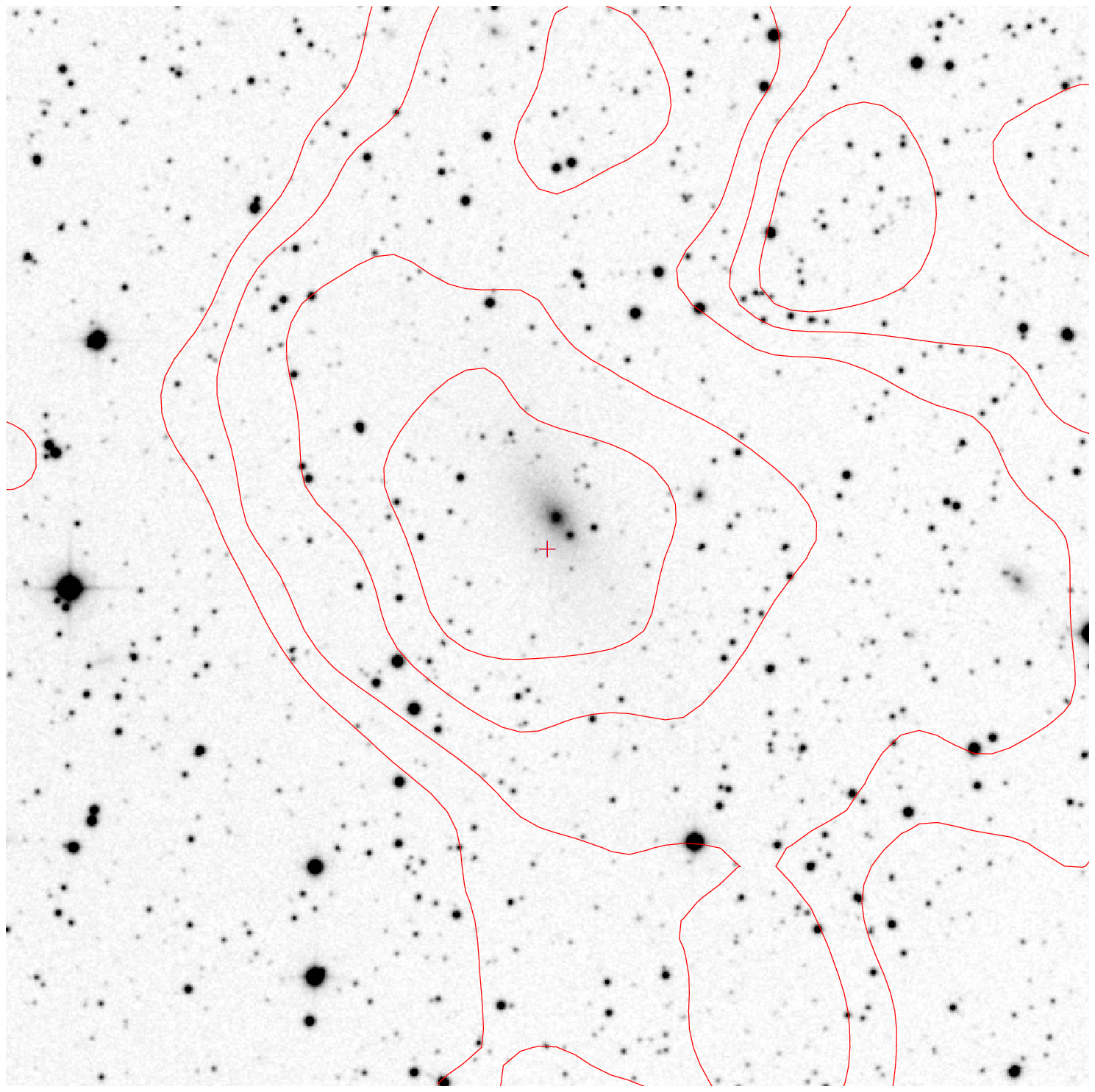}
}
\vskip 1cm 
\hbox{
\hspace{1cm}
   \includegraphics[width=8cm]{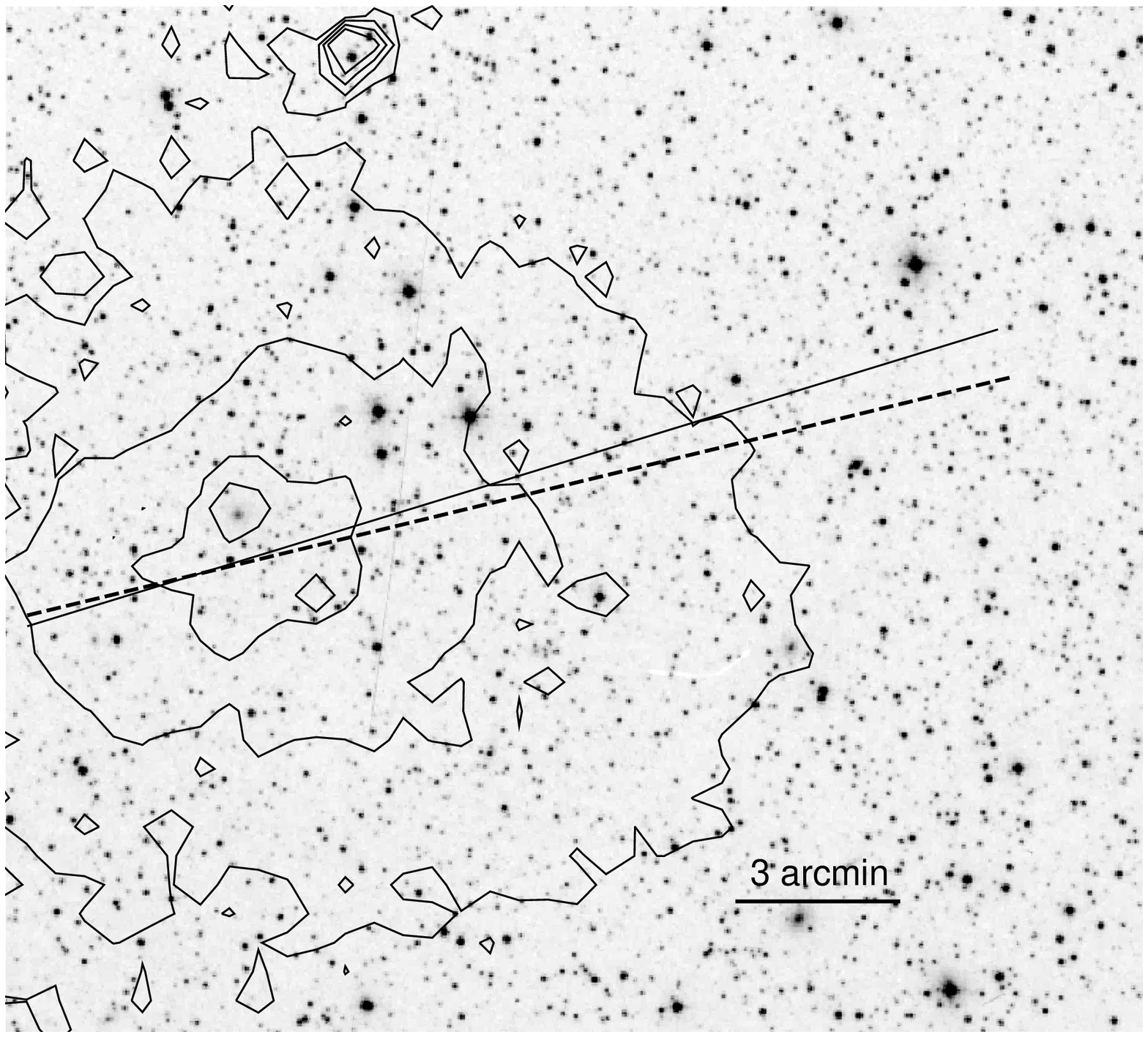}
\hspace{1cm}
   \includegraphics[width=8cm]{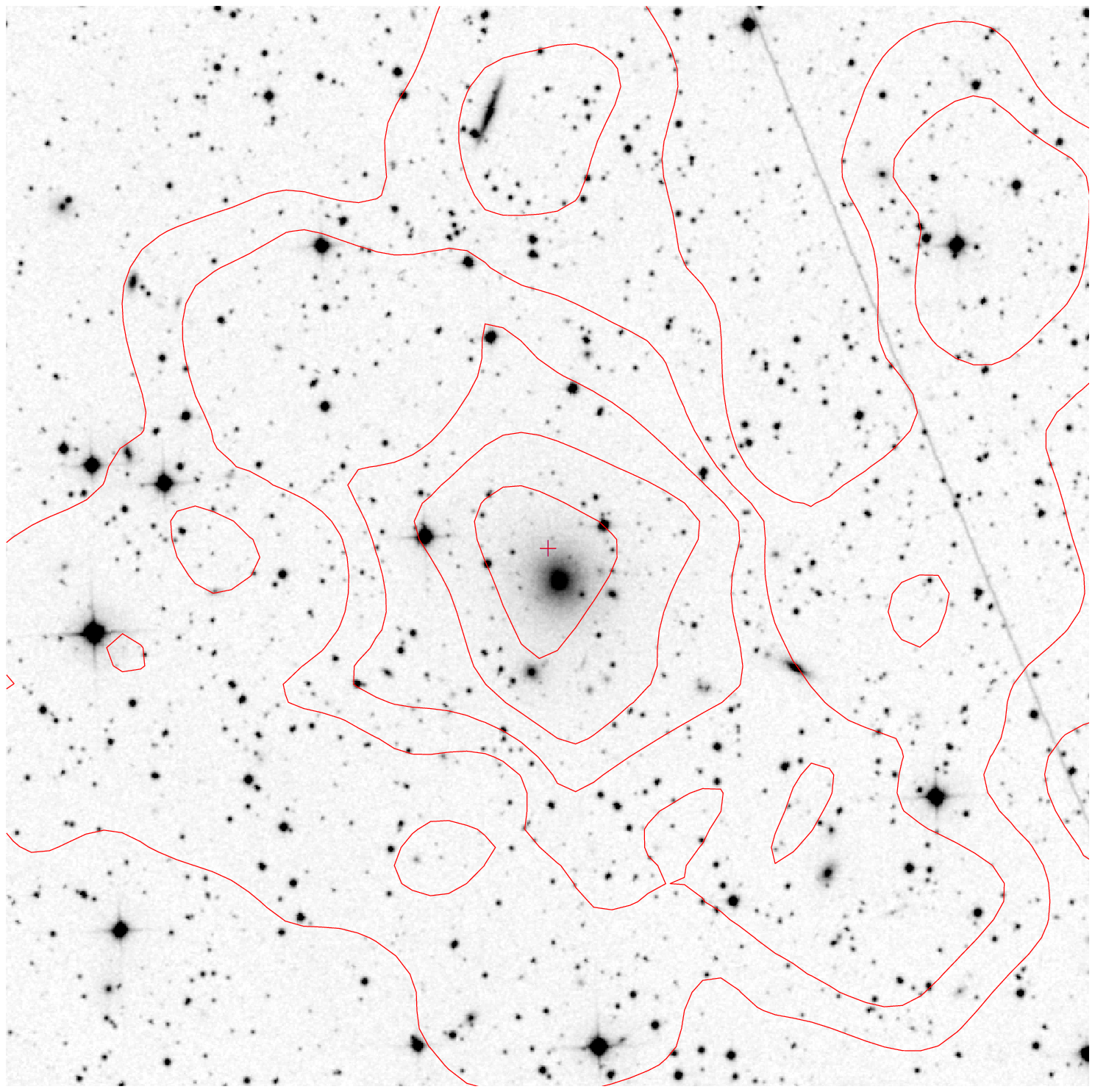}
}
\caption{Images of Perseus-Pisces SC members continued.
{\bf Upper left:}  RXCJ0348.1+4212, MCG +07-08-033, X-ray data from RASS;
{\bf Upper right:} RXCJ0421.8+3607, UGC 03021, X-ray data from RASS;
{\bf Lower left:} RXCJ0450.0+4501, 3C 129, X-ray data from XMM-Newton;
{\bf Lower right:} RXCJ0547.2+5052, UGC 03355, X-ray data from RASS.
}\label{figA1}
\end{figure*}

\subsection{Nature of the X-ray emission in NGC 410}

As we discussed in the main text,  some of the groups in the cluster sample
appear compact. Here we show one of the most compact of these groups in X-rays,
the group associated with NGC410 (RXCJ0110.9+3308). An 
XMM-Newton image of the system is shown in 
Fig. A.1 (lower right). In Fig. A.6 we show the X-ray surface brightness profile 
obtained from the XMM-Newton observation of the group.
The emission is clearly extended, as can be seen by comparison with the point spread 
function (PSF), also shown in the figure. A fit of a beta-model to the profile taking
a convolution with the PSF into account yields a core radius of about 2 kpc.
This is indeed compact compared to the estimated $r_{500}$ of $\sim 400$ kpc.

\citet{Osu2017} studied this system in more detail in X-rays together with
a series of other nearby galaxy groups. They found that the X-ray emission originates
from a thermal plasma with a temperature of about 0.98 keV. Thus, the observed X-ray
emission is clearly consistent with an intragroup medium origin.

\begin{figure}
   \includegraphics[width=\columnwidth]{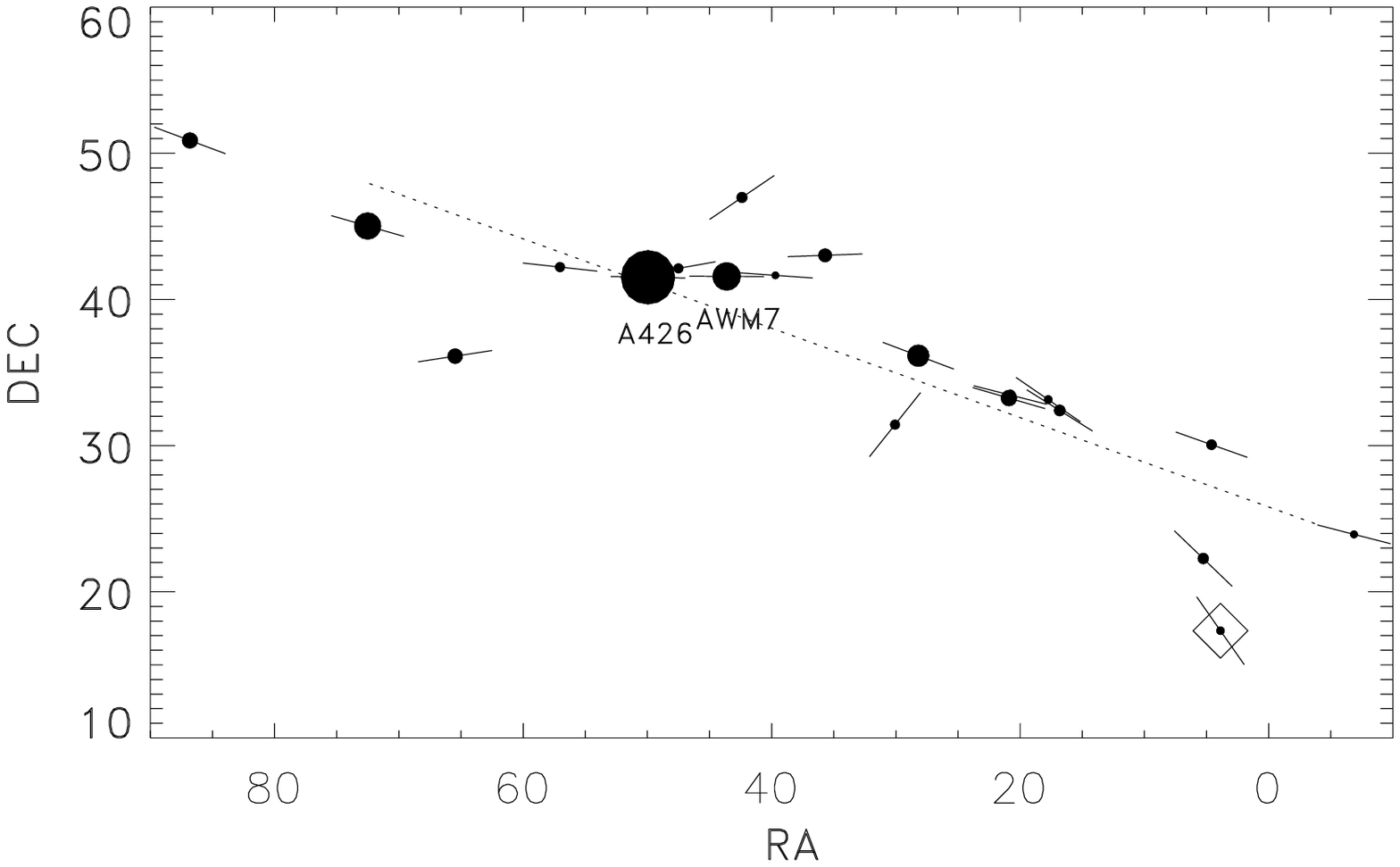}
\caption{Distribution of the members of the Perseus SC 
in equatorial coordinates with 
main axis of the SC (dashed line) and the direction of 
the tidal forces acting on each cluster (short solid lines).
}\label{figB1}
\end{figure}

\begin{figure}
   \includegraphics[width=\columnwidth]{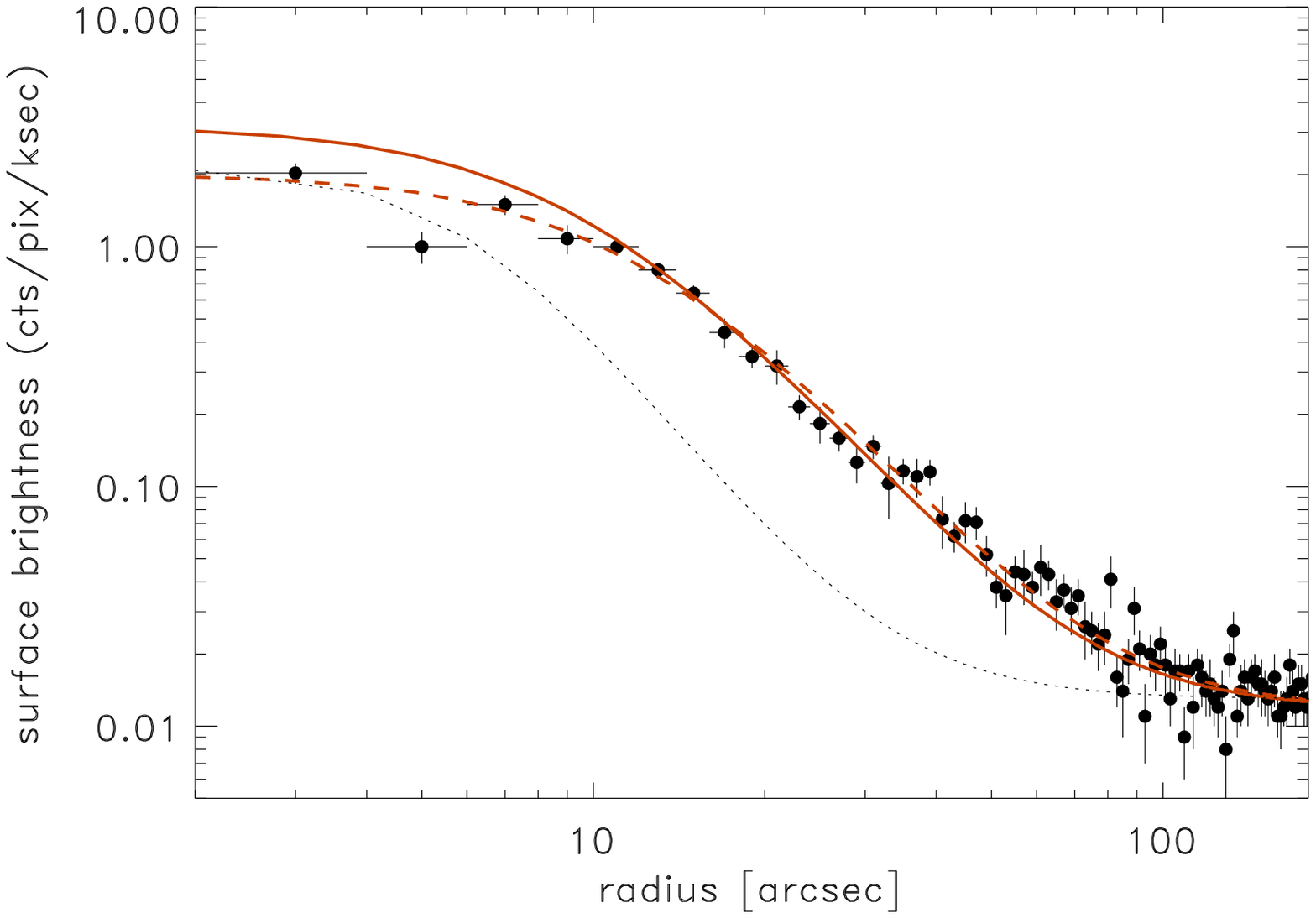}
\caption{X-ray surface brightness profile of NGC 410 traced by the 
MOS detectors of XMM-Newton (data points with error bars). Also
shown is the profile of the point spread function (dotted curve).
The red lines show the best fit of a beta-model, where the dashed
line gives the source profile convolved with the point spread function, which was 
fitted, and the solid line the deconvolved profile.
}\label{figC1}
\end{figure}

\subsection{Images of the members of the Southern Great Wall}

Figures A.7 and A.8 provide images of the seven group and cluster members of 
the Southern Great Wall as derived from the ROSAT All-Sky Survey
overlayed as contours on DSS images.

\begin{figure*}[h]
\hbox{
\hspace{1cm}
   \includegraphics[width=8cm]{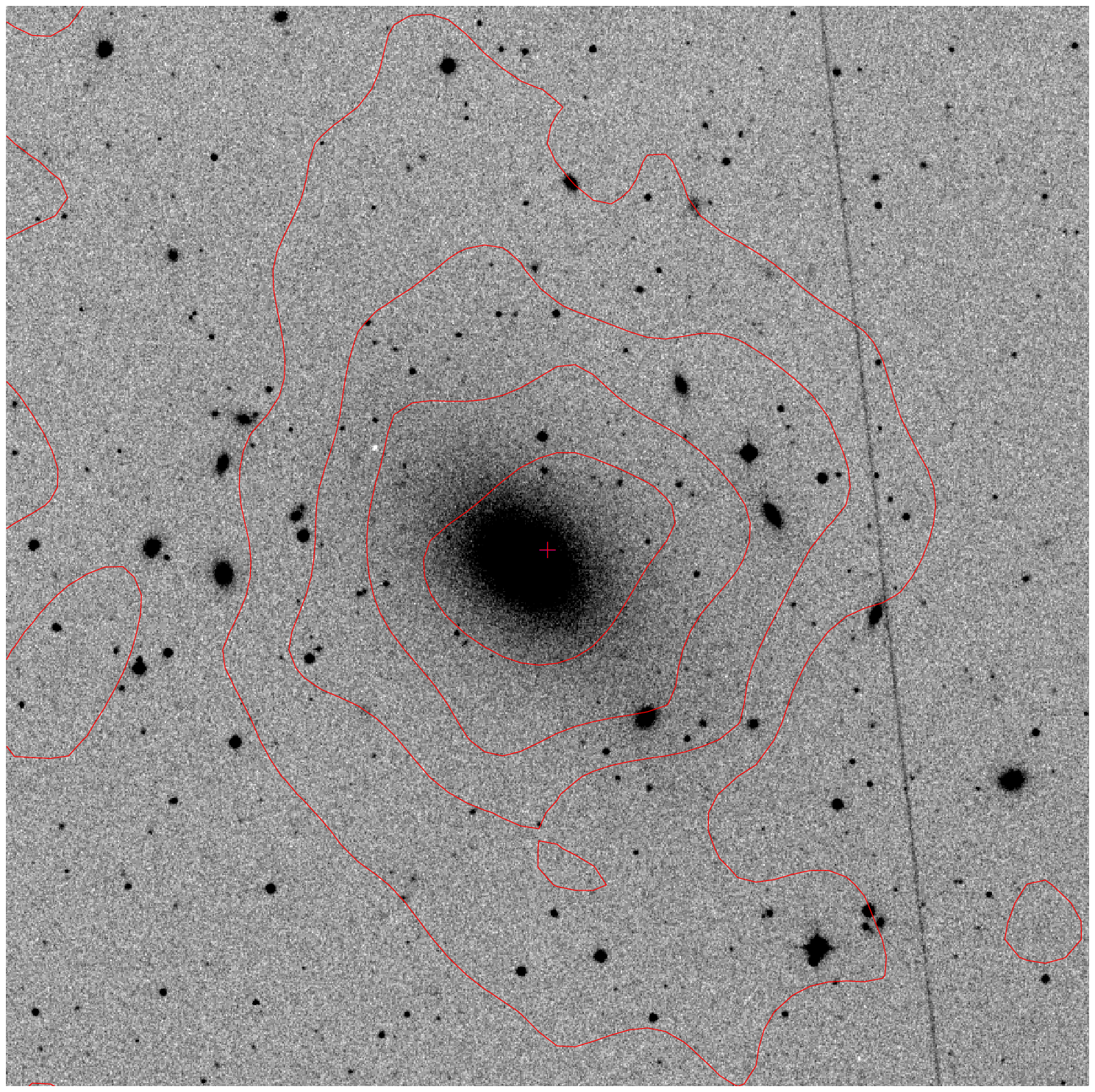}
\hspace{1cm}
   \includegraphics[width=8cm]{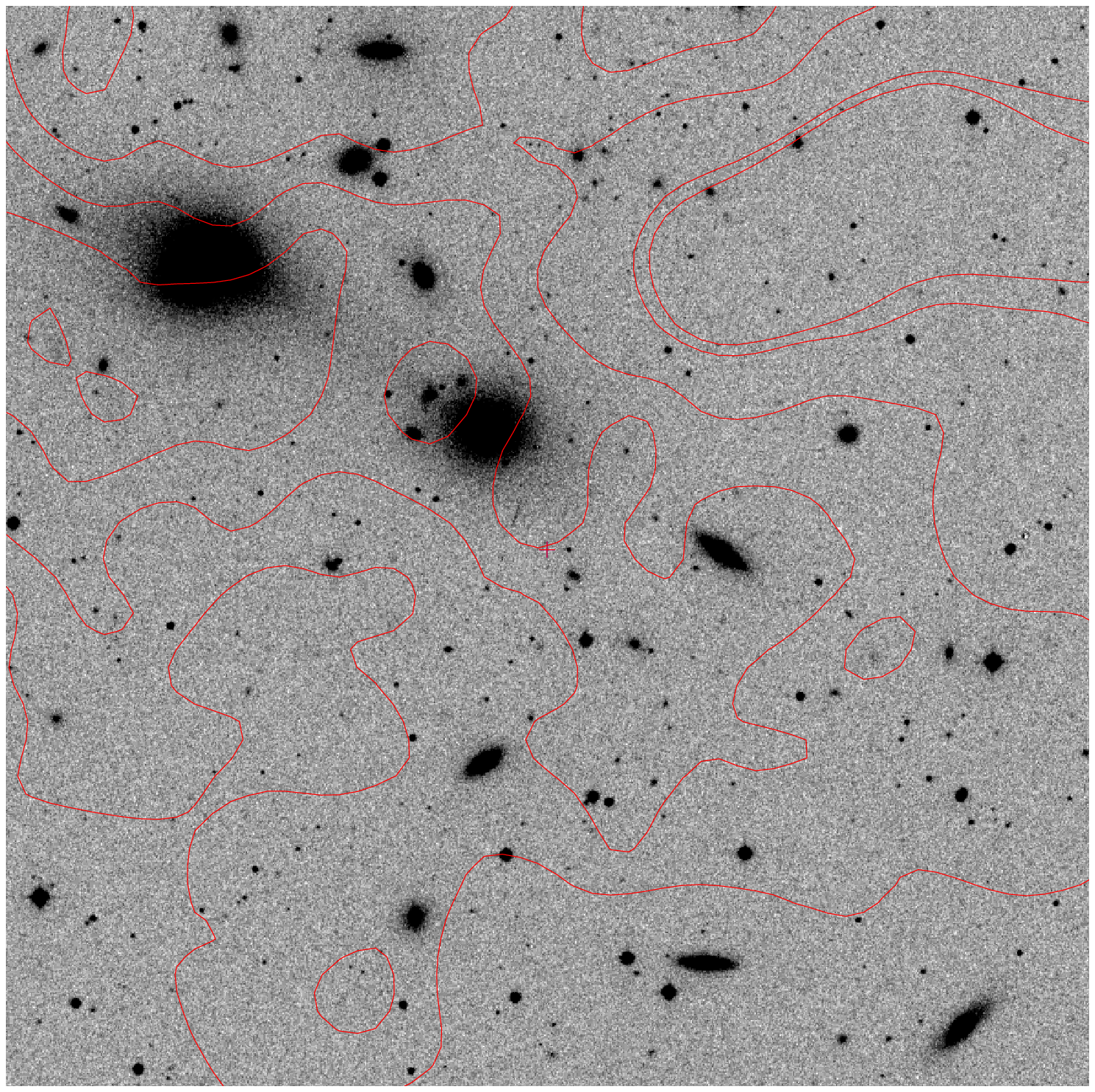}
}
\hbox{
\hspace{1cm}
   \includegraphics[width=8cm]{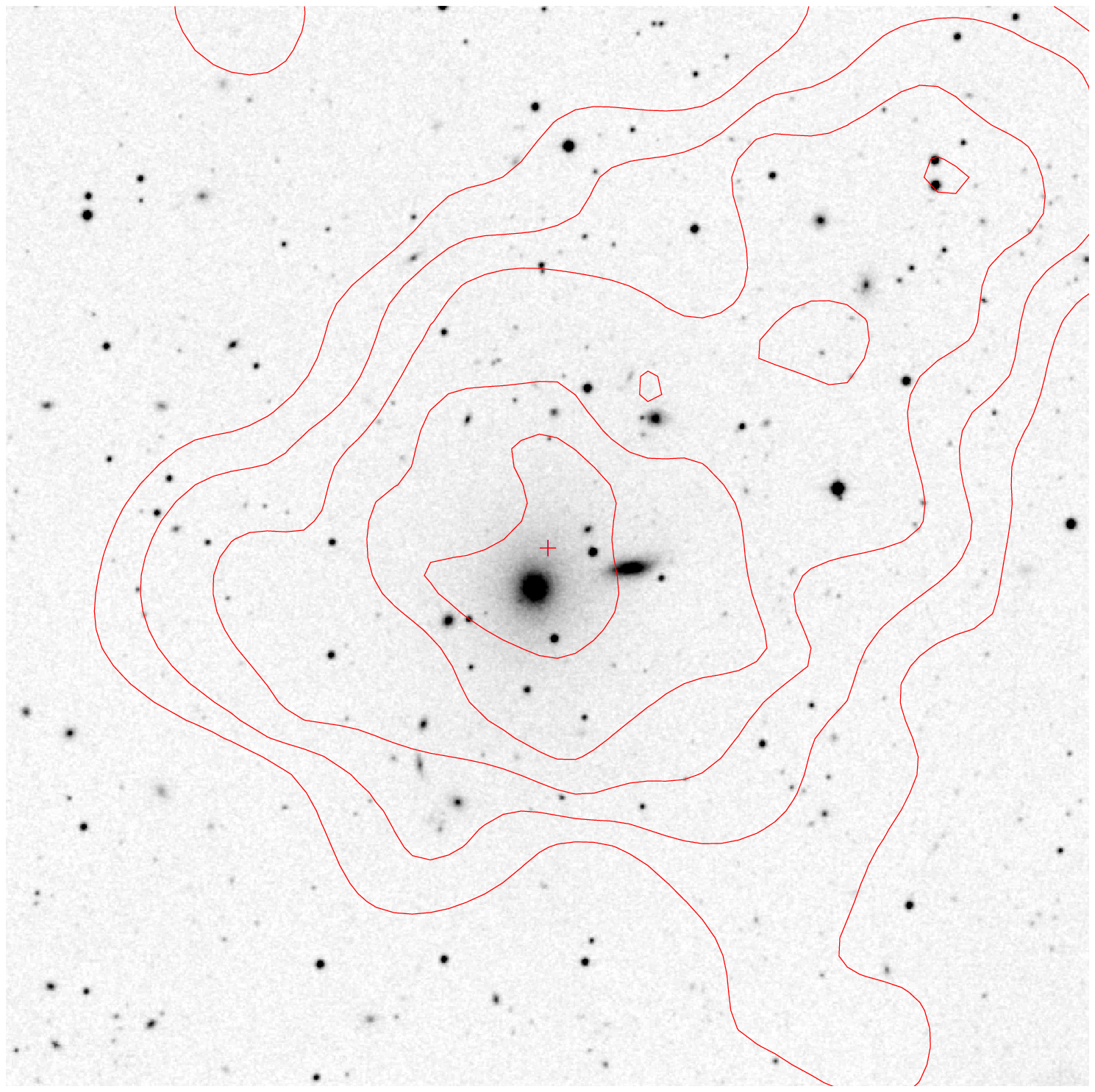}
\hspace{1cm}
   \includegraphics[width=8cm]{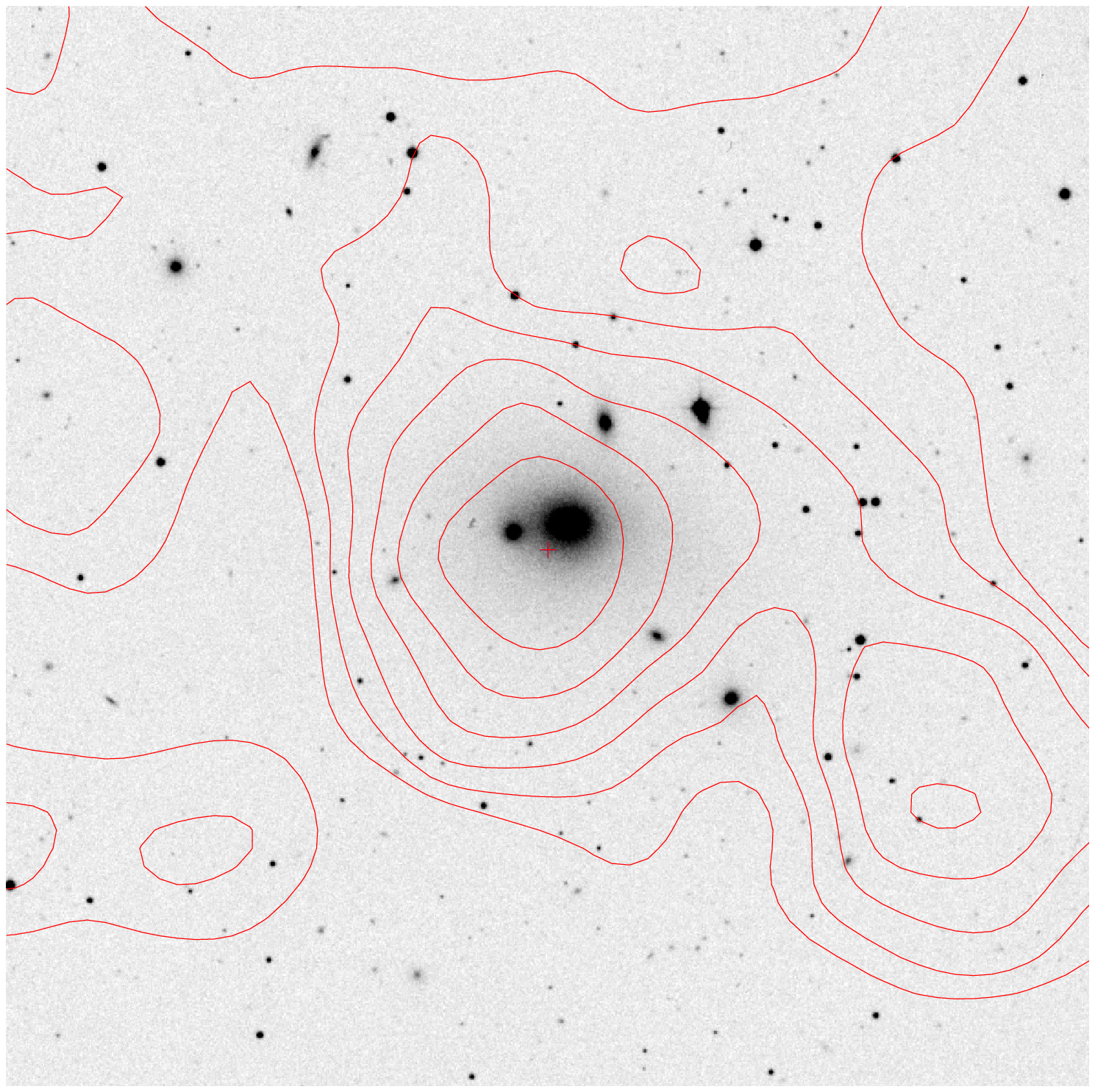}
}
\hbox{
\hspace{1cm}
   \includegraphics[width=8cm]{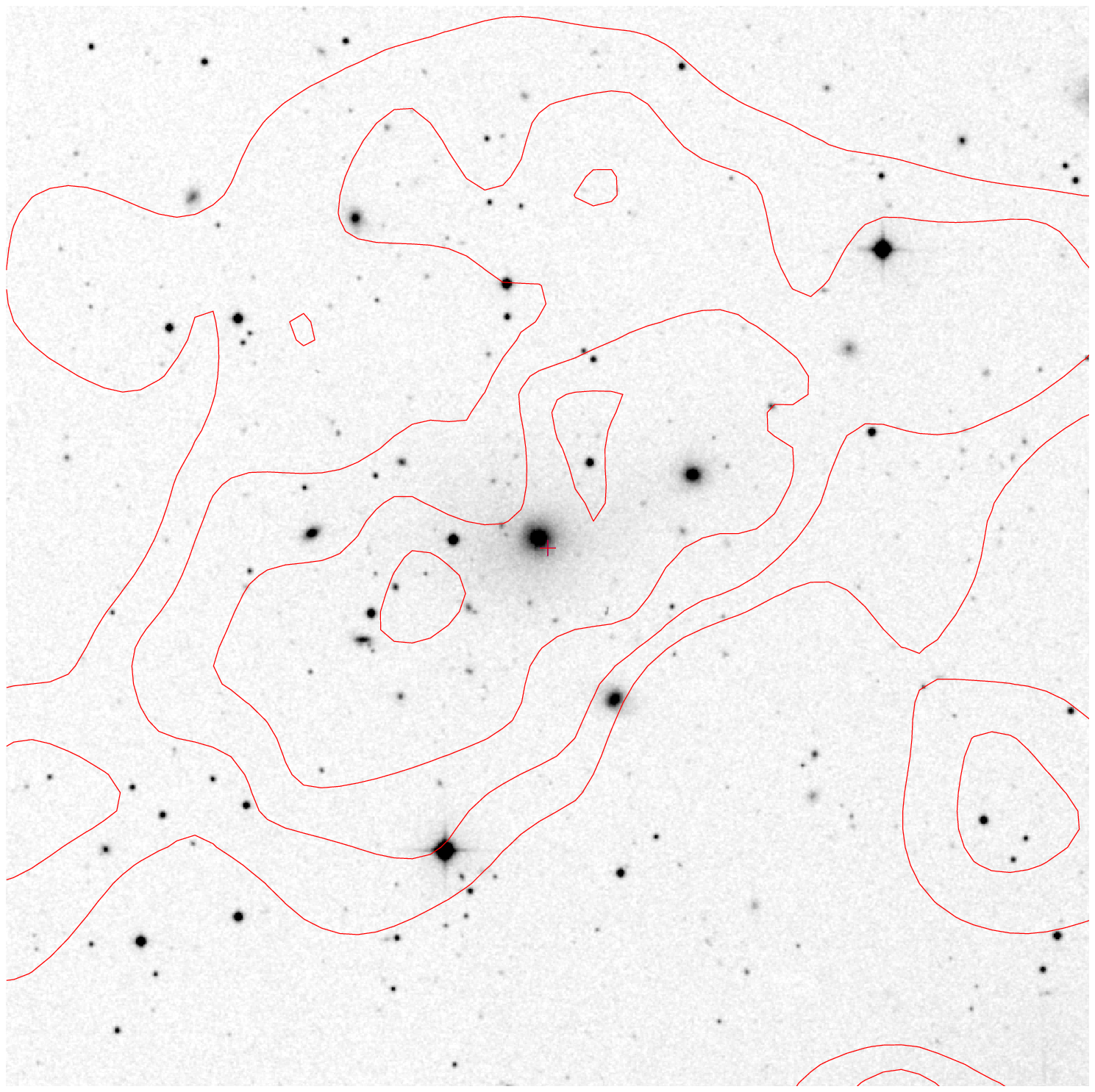}
\hspace{1cm}
   \includegraphics[width=8cm]{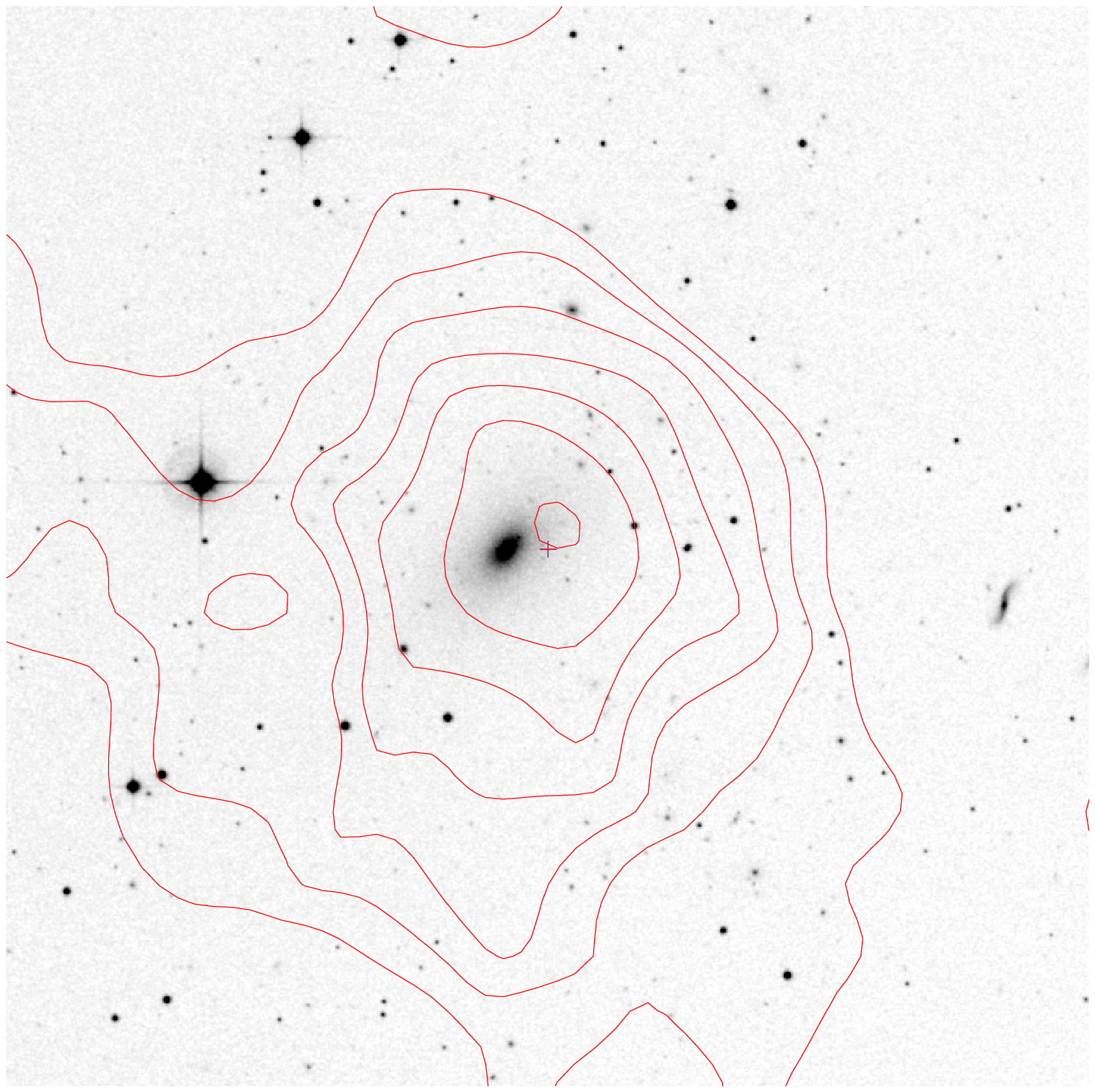}
}
\caption{X-Ray contours overlayed on optical images from the DSS database for the
members of the Southern Great Wall. All X-ray data are  from  RASS.
{\bf Upper left:} RXCJ0125.5+0145, NGC 533; {\bf Upper right:} RXCJ0125.6-0124, A 194;
{\bf Middle left:} RXCJ0149.2+1303, NGC 677;
{\bf Middle right:} RXCJ0156.3+0537, NGC 741;
{\bf Lower left: } RXCJ0231.9+0114, UGC 2005;
{\bf Lower right:} RXCJ0252.8-0116, NGC 1132.
}\label{figA1}
\end{figure*}

\begin{figure}
   \includegraphics[width=\columnwidth]{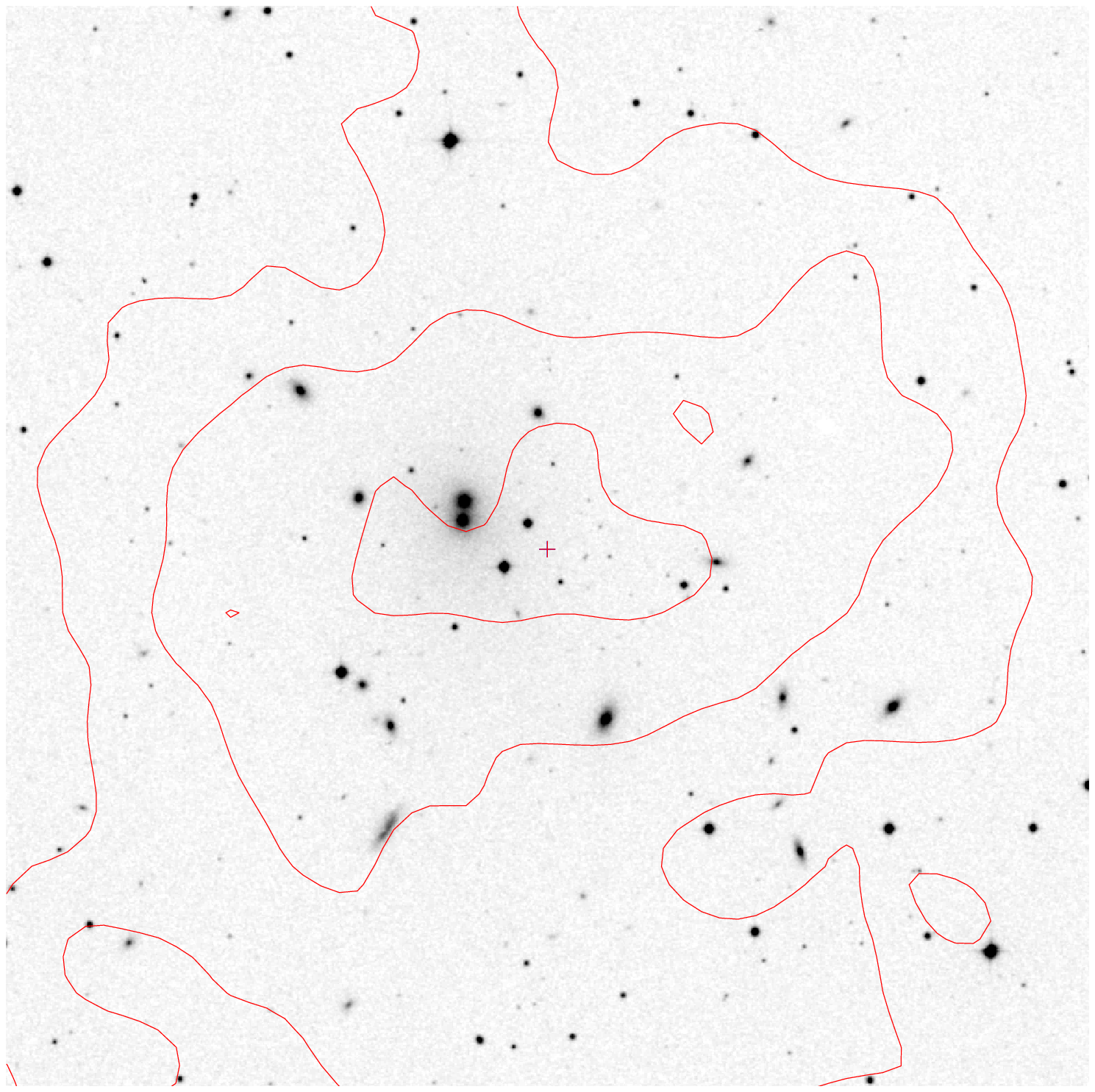}
\caption{Continuation of images of the Southern Great Wall. RXCJ0257.6+0600, 
A 400, with data from RASS.
}\label{figD7}
\end{figure}

\end{document}